\documentclass[sigconf, authorversion=true]{acmart}

\usepackage{hyperref}
\usepackage{comment}
\usepackage{tabularx}
\usepackage{enumitem}
\AtBeginDocument{%
  \providecommand\BibTeX{{%
    \normalfont B\kern-0.5em{\scshape i\kern-0.25em b}\kern-0.8em\TeX}}}

\copyrightyear{2021}
\acmYear{2021}
\setcopyright{none}
\acmConference[EASE 2021]{Evaluation and Assessment in Software Engineering}{June 21--23, 2021}{Trondheim, Norway}
\acmBooktitle{Evaluation and Assessment in Software Engineering (EASE 2021), June 21--23, 2021, Trondheim, Norway}
\acmPrice{}
\acmDOI{10.1145/3463274.3463328}
\acmISBN{978-1-4503-9053-8/21/06}

\begin{document}

%%
%% The "title" command has an optional parameter,
%% allowing the author to define a "short title" to be used in page headers.

\title{Development and Application of Sentiment Analysis Tools in Software Engineering: A Systematic Literature Review}

%%
%% The "author" command and its associated commands are used to define
%% the authors and their affiliations.
%% Of note is the shared affiliation of the first two authors, and the
%% "authornote" and "authornotemark" commands
%% used to denote shared contribution to the research.
\author{Martin Obaidi}
\affiliation{%
  \institution{Leibniz Universität Hannover}
  \institution{Software Engineering Group}
  \country{Germany}
}
\email{martin.obaidi@inf.uni-hannover.de}

\author{Jil Klünder}
\affiliation{%
  \institution{Leibniz Universität Hannover}
  \institution{Software Engineering Group}
  \country{Germany}
}
\email{jil.kluender@inf.uni-hannover.de}

%%
%% By default, the full list of authors will be used in the page
%% headers. Often, this list is too long, and will overlap
%% other information printed in the page headers. This command allows
%% the author to define a more concise list
%% of authors' names for this purpose.
\renewcommand{\shortauthors}{Obaidi and Klünder}

%%
%% The abstract is a short summary of the work to be presented in the
%% article.
\begin{abstract}
Software development is a collaborative task and, hence, involves different persons. Research has shown the relevance of social aspects in the development team for a successful and satisfying project closure. Especially the mood of a team has been proven to be of particular importance. Thus, project managers or project leaders want to be aware of situations in which negative mood is present to allow for interventions. So-called sentiment analysis tools offer a way to determine the mood based on text-based communication. In this paper, we present the results of a systematic literature review of sentiment analysis tools developed for or applied in the context of software engineering. Our results summarize insights from 80 papers with respect to (1) the application domain, (2) the purpose, (3) the used data sets, (4) the approaches for developing sentiment analysis tools and (5) the difficulties researchers face when applying sentiment analysis in the context of software projects. According to our results, sentiment analysis is frequently applied to open-source software projects, and most tools are based on support-vector machines. Despite the frequent use of sentiment analysis in software engineering, there are open issues, e.g., regarding the identification of irony or sarcasm, pointing to future research directions.
\end{abstract}

%%
%% The code below is generated by the tool at http://dl.acm.org/ccs.cfm.
%% Please copy and paste the code instead of the example below.
%%


%%
%% Keywords. The author(s) should pick words that accurately describe
%% the work being presented. Separate the keywords with commas.

%% A "teaser" image appears between the author and affiliation
%% information and the body of the document, and typically spans the
%% page.

%%
%% This command processes the author and affiliation and title
%% information and builds the first part of the formatted document.
\maketitle

\section{Introduction}
\label{sec:introduction}
Due to their growing complexity, software projects are rarely handled by individual developers, but by a team of developers \cite{10.1145/203330.203345,300082}.
Often, these teams are distributed which increases the need for coordination and, hence, for interaction. The HELENA study from 2017 on methods used in software systems development shows that about 60\% of all development teams (n = 1006) are globally distributed \cite{Kuhrmann.2018}. In this context, digital communication tools or channels such as e-mails, Slack\footnote{\url{https://slack.com/}},or JIRA\footnote{\url{https://www.atlassian.com/de/software/jira}} are even more important than in co-located teams where a lot of communication takes place face-to-face \cite{5196929, 10.1145/1882362.1882435}. For meetings, Schneider et al. \cite{SCHNEIDER201859} have shown that negative mood of an individual in the meeting can quickly affect the whole team and afterwards everyone can be demotivated. However, sentiments do not only affect the relationship between two people, but also the productivity, task synchronization and job satisfaction \cite{Graziotin.2014,10.1002/smr.1673,10.1145/2441776.2441812, Guzman.2014}. 

Therefore, detecting bad mood is a goal pursued by many researchers (cf.  \cite{Novielli.,Calefato.2017,Ahmed.2017,Chen.2019,Islam.2017,Islam.2018b}). So-called sentiment analysis tools offer a way to determine the mood based on text-based communication.  
Besides, sentiment analysis tools have also been applied in other application scenarios, including the development of improvement suggestions for codes or recommendations for better software packages and libraries \cite{Murgia.2014,S.Panichella.2015,10.1145/2661685.2661689,7194617}. There is a number of sentiment analysis tools developed and applied in the context of software engineering (SE) \cite{Calefato.2018,Ahmed.2017,Islam.2018}. However, several tools are better suited for different contexts \cite{Novielli.} and have been applied for different reasons and in different scenarios. To get an overview of the state-of-research on sentiment analysis in SE, we conducted a systematic literature review (SLR).

In this paper, we investigate which sentiment analysis tools are used in what domains of the SE and what data sources are used. More concretely, we contribute a list of application scenarios of sentiment analysis tools ranging from applications in academia over applications in open-source projects to applications in industry with their specific purposes, the used data sources, the approaches used for classification and the problems encountered during the development of such tools.

\textit{Outline:} The rest of the paper is structured as follows:
In Section \ref{sec:related}, we present related work. The design of the literature review and the research methodology are explained in Section \ref{sec:review}. The results are presented in Section \ref{sec:results}. In Section \ref{sec:discussion}, we discuss our results, before concluding the paper in Section \ref{sec:conclusion}.

\section{Related Work}
\label{sec:related}
Several authors analyzed the use of specific sentiment analysis tools in SE.
Zhang et al. \cite{9240704} compared sentiment analysis tools like Senti4SD \cite{Calefato.2018} and SentiCR \cite{Ahmed.2017} with each other. In addition, they described models based on the neural network BERT \cite{Devlin.2019}, which were trained with data related to SE such as GitHub\footnote{\url{https://github.com/}} or Stack Overflow\footnote{\url{https://stackoverflow.com/}}. In their replication study, Novielli et al. \cite{Novielli.20.10.2020} explained some sentiment analysis tools (e.g. Senti4SD \cite{Calefato.2018}) in great detail and described the underlying data.

Similarly, other papers compared sentiment analysis tools in their accuracy and described them in terms of their operation \cite{N.Novielli.2018,Novielli.}. Other papers mentioned some tools, too, but only briefly described them without going into details \cite{Biswas.2019,Chen.2019,Islam.2018c}.
In contrast to our work, the authors did not follow a systematic approach to consider the broad range of existing literature and tools, but rather focused on specific papers only. They did not go into detail about why they chose these tools or data and what tools are available.

Nevertheless, there are literature reviews in the field of sentiment analysis that are not related to SE.
Kumar and Jaiswal \cite{Kumar.2020} conducted a SLR with the goal of advancing the understanding of the feasibility, scope, and relevance of studies that apply soft computing techniques for sentiment analysis. They considered tools which used Twitter data and identified research gaps in the field. These gaps include an incessant need to enhance the performance of the sentiment classification tools and the usage of other data sets like Flickr\footnote{\url{https://www.flickr.com/}} or Tumblr\footnote{\url{https://www.tumblr.com/}}.
Abo et al. \cite{Mohamed.2019} conducted a systematic mapping study dealing with sentiment analysis for Arabic texts in social media. Devika et al. \cite{Devika.2016} looked at different approaches to sentiment analysis. Among other approaches such as support-vector machine (SVM), Naive Bayes classifier, they explained rule-based as well as lexicon-based methods.
Maitama et al. \cite{Maitama.2020} performed a systematic mapping study, which contains an examination of aspect-based sentiment analysis tools and an investigation of their approach, technique, diversity and demography.

However, all these SLRs are not related to SE, and the data or tools are not designed to the domain of SE. Consequently, no information about areas or motivation to use the tools in the context of software development is offered.

\section{Literature Review}
\label{sec:review}
In order to gain an overview of the current state of sentiment analysis in the context of SE, we conducted a literature review. In particular, we strive towards reaching the following goal formulated as proposed by Wohlin et al. \cite{Wohlin.2012}:\\

\noindent \fbox{%
\noindent \parbox{\dimexpr\linewidth-2\fboxsep-2\fboxrule}{%
\underline{\textbf{Research Goal: }}\\
\textit{Analyze} existing literature 
\textit{for the purpose of} identifying widely used sentiment analysis methods and tools
\textit{with respect to} different application scenarios in software engineering
\textit{from the point of view of} a researcher
\textit{in the context of} a literature review.}}

\subsection{Research question}
In order to achieve the research goal and to analyze the literature on sentiment analysis in software engineering from different viewpoints, we pose the following research questions:

\noindent \textbf{RQ 1: } \textit{What are the main application scenarios for sentiment analysis in the context of SE?} As a first step, we want to get an overview of the broad area of possible application scenarios in which sentiment analysis is used in the context of software projects.

\noindent \textbf{RQ 2: } \textit{For what purpose is sentiment analysis used in the investigated studies?} Next, we want to analyze the different reasons why sentiment analysis is performed.

\noindent \textbf{RQ 3: } \textit{What data is used as a basis for sentiment analysis?} We want to get an overview of the data used to train or evaluate the tools. This way, we investigate which data is suitable as a basis for sentiment analysis in development teams -- both as training and/or test data.

\noindent \textbf{RQ 4: } \textit{Which approaches are used when developing sentiment analysis tools?} With this question, we gain an overview of good practices in the application of sentiment analysis in SE.  

\noindent \textbf{RQ 5: } \textit{What are the difficulties of these approaches?} Last, we analyze the advantages and disadvantages of existing tools, problems and difficulties, etc. These insights point to future research directions that should be investigated to improve the applicability and the outcome of sentiment analysis tools.

\subsection{Method}
To provide an overview of the development and application of sentiment analysis in the context of SE, we conducted a SLR. Our approach is based on the research process proposed by Kitchenham et al. \cite{KITCHENHAM20097,KitchenhamBA.2007} as well as Petersen et al. \cite{Petersen.2008} and comprises five steps which we describe in the subsequent sections.
\subsubsection{Selection of databases}
\label{subsec:databases}
While some papers are available via several scientific databases, others are not. Therefore, we included a total of five databases in our search to reduce the risk of missing papers which are only available in one database. Our selection comprises databases that are often used in SLRs in the SE domain \cite{8984351, 7929422, kosa.2016, GAROUSI2016106, Klunder.2019, 10.1145/3029387.3029392, 10.1145/3379177.3388907}: 
Science Direct\footnote{\url{https://www.sciencedirect.com/}}, 
IEEE Xplore\footnote{\url{https://ieeexplore.ieee.org/}}, 
ACM Digital Library\footnote{\url{https://dl.acm.org/}}, 
Springer Link\footnote{\url{https://link.springer.com/}}, 
and Google Scholar\footnote{\url{https://scholar.google.com/}}. 
We conducted a comprehensive search as proposed by Petersen et al. \cite{Petersen.2008} in each of these databases in order to reduce biases.

\subsubsection{Definition of the search string}
The search string comprises keywords related to our research questions and basically, the search string uses terms of the two fields of \textit{sentiment analysis} and \textit{SE}. As we focus on SE at its core, we also added related terms such as ``software project'' (which is the typical use case of sentiment analysis) and ``development team'' (which is most often the object whose sentiments are studied). We also extracted synonyms for sentiment analysis like ``opinion mining'' from different papers \cite{8389299,Liu.2012,Liu2012} and considered them in addition to the term ``sentiment analysis''. Finally, we obtained the following search string:\\

\noindent \textit{(``Sentiment analysis'' \textbf{OR} ``text analysis'' \textbf{OR} ``opinion mining'' \textbf{OR} ``emotion AI'')}\\
\textit{\textbf{AND}}\\
\textit{(``software engineering'' \textbf{OR} ``development team'' \textbf{OR} ``software development'' \textbf{OR} ``software project'')}\\

\noindent We adjusted the search string according to the specific syntax of the data bases.

\subsubsection{Definition of inclusion and exclusion criteria}
During the review process, we eliminated studies and publications that cannot contribute to answering our research questions. In order to make this decision more objective, we defined inclusion and exclusion criteria as summarized in Table \ref{tab:inclusion}.
We first applied the exclusion criteria to each of the found publications. In case that none of the exclusion criteria was true for the publication, we decided on the inclusion by considering the inclusion criteria. If the publication fits at least one inclusion criterion, it was included. If a relevant publication appeared more than once (e.g., as a conference paper and as an extended journal publication), we included the most recently published version.

\begin{table}[htb!]
  \caption{Inclusion and exclusion criteria}
  \label{tab:inclusion}
  \begin{tabularx}{\columnwidth}{lX}
    \toprule
   \multicolumn{2}{l}{Inclusion}\\
    \midrule
    1. &The publication presents an approach of the application of sentiment analysis in the context of SE.\\
    2. & The publication presents an approach to creating an sentiment analysis tool/algorithm in the context of SE.\\
    3.& The publication addresses the research questions of this SLR in its goals, hypothesis or applications.\\
    4. &This publication analyzes aspects related with this research.\\
    \toprule
    \multicolumn{2}{l}{Exclusion}\\
    \midrule
    1. &The publication is not written in English.\\
    2. &The publication is not peer-reviewed.\\
    3.& The publication appears repeatedly. In this case, we only considered the latest version.\\
    4.& The publication is not accessible (respectively only accessible only by payment).\\
    5.& The publication has technical content without proven scientific relevance such as invitation papers, editorials, tutorials, keynotes, speeches, white papers, grey literature, dissertations, theses, technical reports, and books.\\
    6.& The publication is a document that is not a full paper or study such as presentations, web postings, web content, citations, brochures, pamphlets, newsletters, or extended abstracts.\\
    \bottomrule
  \end{tabularx}
\end{table}

\subsubsection{Definition of quality assessments}
In order to assess the quality of papers as objectively as possible, we defined quality assessments according to Kitchenham et al. \cite{KITCHENHAM20097,KitchenhamBA.2007}, which can be answered almost objectively. One example is whether a paper provides comprehensible conclusions or not. Depending on whether the criterion is fulfilled completely (=2), partially (=1) or not at all (=0), we assigned the scoring to the paper. If a criterion was not applicable, we did not take it into account. If the average of all applicable quality attributes was less than 1, we considered the paper as having insufficient quality and removed it from further analyses. This happened five times.
\subsubsection{Execution}
\begin{figure*}[htb!]
  \centering
  \includegraphics[width=0.60\textwidth, keepaspectratio]{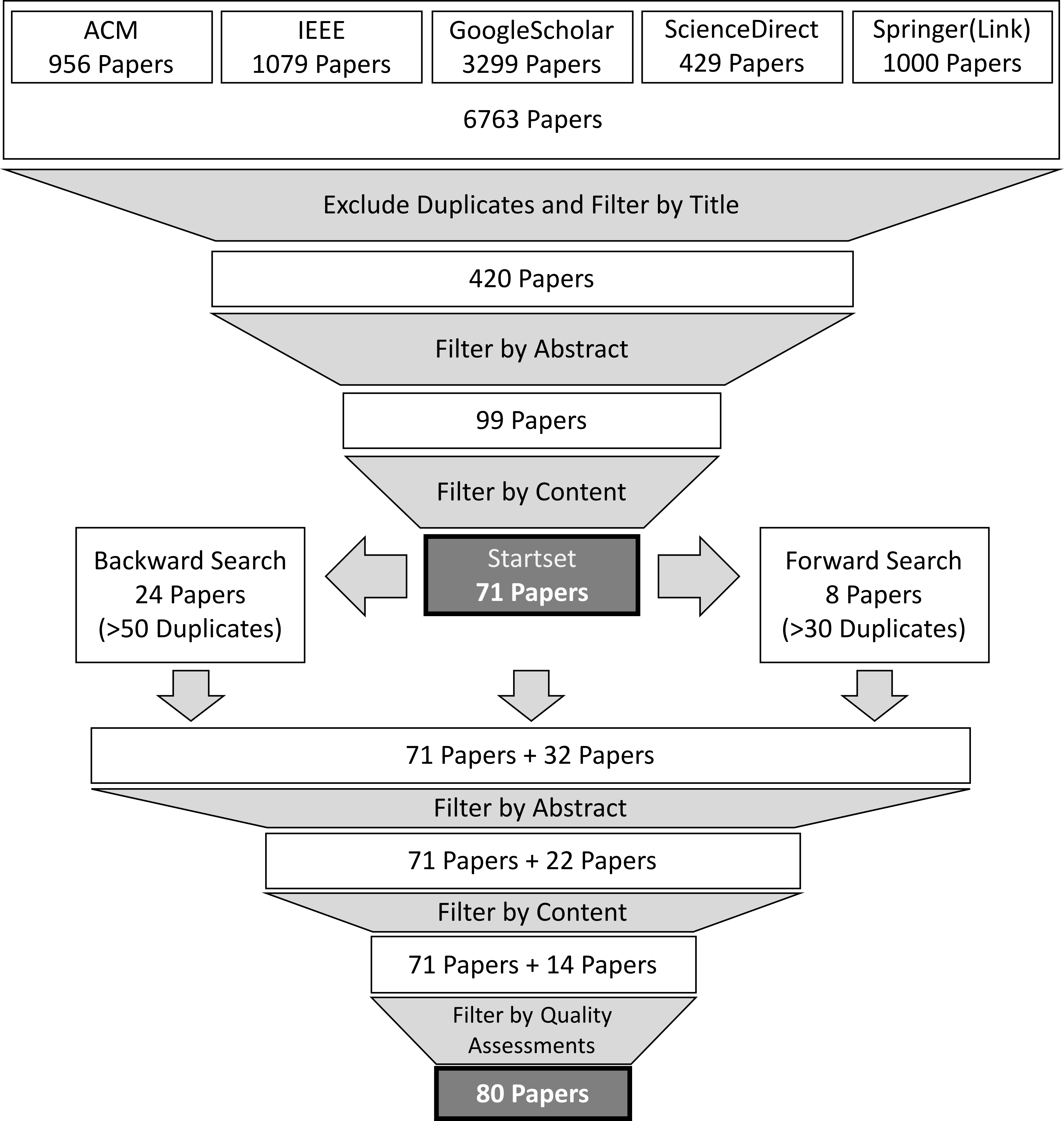}
  \caption{Search process and filtering steps}
  \label{fig:overview}
\end{figure*}

An overview of the execution can be seen in Figure \ref{fig:overview}.
First, we searched the five databases mentioned in subsection \ref{subsec:databases} using the search string and selected probably relevant papers based on their title from a total of 6763 publications. After removing duplicates and filtering the papers based on the titles, we applied the inclusion and exclusion criteria to the remaining 420 papers' abstracts. Ninety-nine papers remained of which we scanned the full text, resulting in 71 papers used as the startset for the forward and backward search. This snowball principle was performed according to Wohlin \cite{Wohlin.2014}. The backward search resulted in 24 new papers based on the title, and the forward search resulted in 8 papers. We again scanned the 32 new papers based on abstract and full text, leading to 14 more papers considered as relevant.

In total, we identified 85 papers as relevant: 71 in the initial search and 14 in one forward and backward search. The number of new papers after one snowball iteration was small. Therefore, we refrained from repeating the snowball principle after having examined the first papers and could not find any relevant new papers based on the title, except duplicates. 

We ended our search for papers at the end of December 2020.

The quality metrics were applied to the 85 papers and five were excluded due to not passing the quality criterion. Some of them did not define research questions or discuss results, or the technologies used were not described. All of them had an average score less than 1 according to the quality assessments. 

In this execution phase, we listed each relevant paper in an Excel spreadsheet \cite{martin_obaidi_2021_4726651}. We included relevant parts of the paper, and based on this data, we clustered the papers and created categories for the research questions. 

\subsubsection{Threats to Validity}
The outcome of our literature review is biased by different factors. We cannot assume that we found all publications in the context of our research nor that our results are complete. In fact, since we found few papers that are not accessible, it is not unlikely that there are relevant papers that we could not include in our results.

In the following, we present threats to validity according to the different steps of the literature review.

\textit{Selection of databases}: In order to find as many relevant papers as possible, a high number of used databases is an advantage. Some publications are listed in more than one database while others are not (construct validity). For our SLR, we used five scientific databases that are known to be relevant in the area of SE and have been used by several other SLRs in SE (e.g., \cite{8984351, 7929422, kosa.2016, GAROUSI2016106, Klunder.2019, 10.1145/3029387.3029392, 10.1145/3379177.3388907}).

\textit{Definition of the search string}: We composed the search string from the two areas \textit{SE} and \textit{sentiment analysis}. We used synonyms for \textit{sentiment analysis}, as well as several words from the field of SE. Nevertheless, there may be other synonyms or related terms of which we are currently unaware (construct validity).
Although the search string affects the outcome of the study, we are confident that the accuracy of the results found based on our search string suffices to answer our research questions and to achieve our  goal of getting an overview on existing literature.

\textit{Definition of the inclusion and exclusion criteria}: Certain publications can be included or excluded based on various characteristics. For example, we excluded papers that have not been peer-reviewed. In order to be as objective as possible and to reduce bias due to subjective decisions (internal validity), we formulated criteria for the inclusion and exclusion of a paper based on Kitchenham et al. \cite{KITCHENHAM20097,KitchenhamBA.2007}. Some of the criteria like the accessibility of a paper are purely objective, while the evaluation of the content regarding the relevance to the study is still somewhat subjective. In case of doubts, we included the paper (and possibly excluded it later after having reviewed the abstract or content) and then decided how to proceed with the paper.

\textit{Definition of quality assessment}: In addition to the inclusion and exclusion criteria, we set up quality assessments according to Kitchenham et al. \cite{KITCHENHAM20097,KitchenhamBA.2007}. These assessments provide an objective framework to assess the quality of the paper and look at whether certain standards have been met, rather than arbitrarily including or excluding a paper. The lower limit for filtering out was chosen as 1, which means that all applicable quality criteria are on average at least partially fulfilled. 

\textit{Execution}: The study was largely conducted by one researcher, but the whole research process was reviewed by other researchers. Therefore, some researcher bias is present (internal validity), but we are confident that this influence only plays a minor role.

\textit{Results}: Due to the design of the study, we cannot guarantee that we found all papers relevant to the research objective. In addition, a repeated application of the snowball principle would have found other possible relevant papers. However, we are confident that the papers we found are sufficient to answer the research questions.

\section{Results}
\label{sec:results}
Our literature review revealed 80 publications dealing with the topic of sentiment analysis in the context of SE. These publications are available online as raw data set \cite{martin_obaidi_2021_4726651}.

\subsection{Application Domain}
During the classification phase based on the research question, we identified three application domains in which sentiment analysis was developed or applied: (1) open-source software (OSS) projects, (2) industry, and (3) academia. In case that the application domain was not explicitly named, we selected the domain based on the data sets used or the context of the usage of sentiment analysis. If, for example, GitHub was used, we aligned the publication to OSS domain. If data sets like app reviews were used, we chose the industrial domain.

Of the $n=80$ papers, 59 belong to the domain of OSS, 17 to the domain of industry and 6 to the domain of academia. An example of application in an industrial context is a sentiment analysis on developers chat communication data at Amazon MTurk \cite{10.1145/3392877}. University applications include sentiment analysis among students in software projects organized by universities \cite{Gkontzis.2017,Guzman.2013b,Guzman.2013}.\\

\noindent \fbox{%
\noindent \parbox{\dimexpr\linewidth-2\fboxsep-2\fboxrule}{%
\underline{\textbf{Finding:}} Most of the papers on sentiment analysis in SE are based on open-source projects. Less then 1/3 of the papers either considers industrial projects or the academia.}}

%\noindent \fbox{%
%\noindent \parbox{\dimexpr\linewidth-2\fboxsep-2\fboxrule}{%
%\underline{\textbf{Finding: }}\\
%\noindent Most of the papers on sentiment analysis in SE are based on open-source projects. Less then 1/3 of the papers either considers industrial %projects or the academia.}}

%\noindent \fbox{%
%    \begin{minipage}{\dimexpr\linewidth-2\fboxsep-2\fboxrule}%
%    \begin{minipage}[b]{0.9\linewidth}
%        \underline{\textbf{Finding: }}\\
%        \noindent Most of the papers on sentiment analysis in SE are based on open-source projects. Less then 1/3 of the papers either considers %industrial projects or the academia.
%    \end{minipage}
%
%    \end{minipage}}
    
%\subsection{Purpose of Using Sentiment Analysis}
\subsection{Purpose}
\label{sec:rq2}
When investigating the motivation of the papers, we distinguish between three types of papers: (1) development, (2) comparison, and (3) application of sentiment analysis tools. These categories can be seen as primary motivation that can be refined further.
The first type, \textit{development}, consists of papers that aim to develop a sentiment analysis tool. The second type compares already existing tools with each other, while the third type focuses on their application. There are papers that developed a new tool and then compared it to existing tools. Since all these papers had the primary goal of developing a new tool, we assigned them to the type \textit{development}.

Of the $n=80$ papers, 16 belong to development, 12 to comparison and the remaining 52 to application. Details are summarized in Table \ref{tab:rq2}.
The papers of the type \textit{development} have the main purpose of developing a procedure for sentiment analysis in order to examine certain data in the context of SE, such as developer communication (e.g. \cite{Calefato.2018}), for sentiment. From $n=12$ papers of the type \textit{comparison}, 9 strive to find the best sentiment analysis tool among several tools. Three want to compare the allocation of sentiments from existing sentiment analysis tools with the allocation from humans. From $n=52$ papers of the type \textit{application}, 35 want to find correlations between sentiments an specific values, whereas 18 analyze social aspects of developers. Seven papers want to measure a specific value like subjective usability (e.g. \cite{ElHalees.2014}) or the marketability of an open-source app (e.g. \cite{Nayebi.2017}). Five papers use sentiment analysis tools, among other methods, to predict special values like the performance of a teacher based on feedback from students (e.g. \cite{.2015,Aung.2017}).

\begin{table}[htbp!]
\caption{Motivations of the papers in detail}
\label{tab:rq2}
\begin{tabular}{lllllll}
\toprule
Type        & \rotatebox[origin=l]{90}{Find best tool} & \rotatebox[origin=l]{90}{Tools vs. human} & \rotatebox[origin=l]{90}{Correlations}  & \rotatebox[origin=l]{90}{Social aspects} & \rotatebox[origin=l]{90}{Values measurements} &  \rotatebox[origin=l]{90}{Values predictions} \\\midrule
Comparison  & 9    & 3      & 0              & 0           & 0       & 0\\
Application & 0    & 0     & 35              & 18           & 7       & 5\\
\bottomrule
\end{tabular}
\end{table}

\noindent \fbox{%
\noindent \parbox{\dimexpr\linewidth-2\fboxsep-2\fboxrule}{%
\underline{\textbf{Finding:}} There are three types of papers, based on their main purpose: Development of sentiment analysis tools, comparison and application of them. Most of the papers belong to the application type, whereas 35\% of the papers either belong to development or comparison.}}

\subsection{Used data sources}
Next, we analyzed which data is used for training or evaluation for sentiment analysis tools. We identified a total of 48 data sources. Most of them occur only one or two times, because the data used in the application domain are often unique (such as a specific data set of a student software project). Therefore, for the sake of clarity, we listed all data sets in Table \ref{tab:rq3} that occurred at least three times in total. Data sets that occurred less then three times are summarized as ``other''. 

\begin{table}[htbp!]
\caption{Overview of the used data}
\label{tab:rq3}
\begin{tabular}{llllllll}
\toprule
Type        & \rotatebox[origin=l]{90}{JIRA} & \rotatebox[origin=l]{90}{GitHub} & \rotatebox[origin=l]{90}{Stack Overflow} & \rotatebox[origin=l]{90}{App reviews} & \rotatebox[origin=l]{90}{Twitter} & \rotatebox[origin=l]{90}{Code reviews} & \rotatebox[origin=l]{90}{Other}\\\midrule
Development & 6    & 3      & 9              & 0           & 2       & 1            & 3\\
Comparison  & 9    & 5      & 7              & 1           & 0       & 2            & 5\\
Application & 6    & 12     & 4              & 7           & 2       & 0            & 38\\ \bottomrule
Total       & 21   & 20     & 20             & 8           & 4       & 3            & 46\\
\bottomrule
\end{tabular}
\end{table}

The papers from the development and comparison categories use the same data source for training and testing an algorithm. Only Chen et al. \cite{Chen.2019} use GitHub emojis for fine-tuning an existing neural network, but not for testing. It is notable that from the $n=80$ papers, the data sources \textit{JIRA}, \textit{GitHub} and \textit{Stack Overflow} are the most represented with 21, 20 and 20 uses respectively. App reviews were used 8 times, while all other data sources were used at most 4 times. In the application domain alone, from $n=52$ papers, mostly unique (37 out of 38) data sources are used to apply a sentiment analysis tool to them. Other data sources include chat data from Amazon MTurk \cite{10.1145/3392877}, Amazon product review \cite{Fang.2015}, android bug reports \cite{Umer.2020}, or support tickets from IBM \cite{Werner.2018}.\\

\noindent \fbox{%
\noindent \parbox{\dimexpr\linewidth-2\fboxsep-2\fboxrule}{%
\underline{\textbf{Finding:}} The results show that there are 48 different data source. Six of them are used at least three times. JIRA, GitHub and Stack Overflow are used most frequently for training and testing sentiment analysis algorithms.}}

\subsection{Approaches for developing or using sentiment analysis}
In line with the results presented in subsection \ref{sec:rq2}, we again distinguish between \textit{development} and \textit{application} of the tools when looking at the algorithms the sentiment analysis is based on. For example, when a paper is about the development of a sentiment analysis tool, we listed which machine learning approach this tool uses, such as SVM or Bayes. If an existing tool like SentiStrength \cite{Thelwall.2010,Thelwall.2012} is used, we listed this specific tool. Some papers did both developing and comparing their tool with other existing tools.

An overview of approaches being used during developing can be seen in Table \ref{tab:rq4-1}. Because only existing tools were compared in the ``comparison'' category, we did not list the respective papers. For the sake of clarity, we summed up all approaches that appeared less than three times in ``other''.

\begin{table}[htb!]
\caption{Overview of approaches used during developing sentiment analysis tools}
\label{tab:rq4-1}
\begin{tabular}{llllllllll}
\toprule
Type    & \rotatebox[origin=l]{90}{Bayes} & \rotatebox[origin=l]{90}{SVM} & \rotatebox[origin=l]{90}{Neural network} & \rotatebox[origin=l]{90}{Random forest} & \rotatebox[origin=l]{90}{Logistic regression} & \rotatebox[origin=l]{90}{Decision tree} & \rotatebox[origin=l]{90}{Lexicon/Heuristic} & \rotatebox[origin=l]{90}{Gradient boosting} & \rotatebox[origin=l]{90}{Other}\\ \midrule
Development & 5     & 7   & 4              & 3             & 3                   & 2             & 3                 & 3 & 6\\ 
Application & 6     & 4   & 3              & 2             & 1                   & 1             & 0                 & 0 & 1\\ \bottomrule
Total       & 11    & 11  & 7              & 5             & 4                   & 3             & 3                 & 3  &  7\\ \bottomrule   
\end{tabular}
\end{table}

From the remaining $n=68$ papers providing information on the used approaches, we found a total of 15 different machine learning approaches used for evaluations. With 11 times, SVM and different kind of Bayes classifiers (e.g. Naive Bayes) were used most frequently. However, also other established methods like random forest, logistic regression or AdaBoost were used, too. Three of the 15 machine learning methods were used only one time: Sequential minimum optimization \cite{Cagnoni.2020}, bootstrap aggregating \cite{Ding.2018} and pattern-based approach \cite{Lin.2019}.

Some papers have also compared several approaches. From the $n=68$ papers, 9 compared different machine learning approaches and chose the best performing one. In these comparisons, SVM won the comparison with three times the most. Gradient boosting is second and won two times. Bayes, random forest, logistic regression and neural network won only one time.\\

Table \ref{tab:rq4-2} gives an overview about the application of existing sentiment analysis tools. In the $n=80$ papers, we found a total of 28 tools. For the sake of clarity, we summed up all tools that appeared less than three times in ``other''.

\begin{table}[htb!]
\caption{Overview of sentiment analysis tools applications. Tools marked with * are specifically designed for the SE domain)}
\label{tab:rq4-2}
\begin{tabular}{lllllllllll}
\toprule
Type            & \rotatebox[origin=l]{90}{SentiStrength} & \rotatebox[origin=l]{90}{NLTK} & \rotatebox[origin=l]{90}{SentiStrength-SE*} & \rotatebox[origin=l]{90}{Senti4SD*} & \rotatebox[origin=l]{90}{CoreNLP} & \rotatebox[origin=l]{90}{SentiCR*} & \rotatebox[origin=l]{90}{Vader} & \rotatebox[origin=l]{90}{Alchemy} & \rotatebox[origin=l]{90}{WordNet} & \rotatebox[origin=l]{90}{Other}\\ \midrule
Development     & 1             & 1    & 1                & 1        & 1        & 1       & 0     & 0       & 0      &  1\\
Comparison      & 8             & 6    & 5                & 5        & 4        & 5       & 1     & 3       & 1      & 12\\
Application     & 20            & 6    & 4                & 3        & 1        & 0       & 3     & 0       & 2      & 15 \\ \bottomrule
Total           & 29            & 13   & 10               & 9        & 6        & 6       & 4     & 3       & 3      & 28 \\ \bottomrule
\end{tabular}
\end{table}

The results show that three tools stand out, which were used at least 10 times. With 29 times, SentiStrength \cite{Thelwall.2010,Thelwall.2012} is the most used sentiment analysis tool by far. The second most used tool is SentiStrenght-SE \cite{Islam.2017,Islam.2018}, which is an adaptation of SentiStrength to the domain of SE. It was used 10 times. NLTK \cite{Loper.17.05.2002}, which is a  natural language toolkit and handles sentiment analysis, was used 13 times. All other sentiment analysis tools were used less than 10 times, often only 1-2 times total. These include, for example, the neural network BERT \cite{Devlin.2019}, the BERT-based model RoBERTa \cite{liu2019roberta}, the open source machine learning software WEKA \cite{396988} or the lexicon-based tool DEVA \cite{Islam.2018b}.\\

\noindent \fbox{%
\noindent \parbox{\dimexpr\linewidth-2\fboxsep-2\fboxrule}{%
\underline{\textbf{Finding:}} The results show that 28 different existing sentiment analysis tools were used in the $n=80$ papers, with SentiStrength standing out. Concerning the different machine learning methods there are 15, which were used for evaluation in the $n=80$ papers. SVM and Bayes stand out here. The authors often chose SVM because of its good performance.}}

\subsection{Difficulties}
\label{sec:difficulties}
Among all papers, some difficulties regarding the field of sentiment analysis in SE were mentioned frequently. Table \ref{tab:rq4-3} shows an overview of the results of the mentioned problems.

\begin{table}[!htb]
\caption{Overview of mentioned problems regarding sentiment analysis in SE}
\label{tab:rq4-3}
\begin{tabular}{llllllll}
\toprule
Type        & \rotatebox[origin=l]{90}{Adaption to the domain of SE} & \rotatebox[origin=l]{90}{Sarcasm/Irony} & \rotatebox[origin=l]{90}{Subjectivity of manual labeling} & \rotatebox[origin=l]{90}{Small amount of data} & \rotatebox[origin=l]{90}{Cross-platform performance}\\ \midrule
Development & 9           & 3               & 0             & 4                    & 1\\ 
Comparison  & 5           & 3               & 4            & 1                    & 1\\ 
Application & 12          & 5               & 6            & 1                    & 2\\ \bottomrule
Total       & 26          & 11              & 10           & 6                    & 4   \\ \bottomrule
\end{tabular}
\end{table}

Of the $n=80$ papers, 26 mentioned the problem of the lack or scarcity of adaptations of existing sentiment analysis tools to the domain of SE (e.g. \cite{Ahmed.2017,Calefato.2018,Chen.2019,Ding.2018,Imtiaz.2018}). However, other problems like sarcasm/irony handling (11) or the subjectivity of manual labeling of data (10) were also mentioned. In addition, there are also investigations of how different sentiment analysis tools perform, when they are trained in a cross-platform setting \cite{Novielli.}. Novielli et al. \cite{Novielli.} mention that the tools trained with one data set often performed poorly when they are tested with a different data set \cite{Novielli.}.\\

\noindent \fbox{%
\noindent \parbox{\dimexpr\linewidth-2\fboxsep-2\fboxrule}{%
\underline{\textbf{Finding:}} There are some difficulties regarding sentiment analysis in SE, which are mainly related to subjective data labels, too much customization for specific data sets and no/partial adaptation to the domain of SE.}}

\section{Discussion}
\label{sec:discussion}

\subsection{Answer to research questions}
\textit{RQ 1: What are the main application scenarios for sentiment analysis in the context of SE?}\\
Our results show that there are three application domains: Open-source software domain, industry  and academia. The $n=80$ papers were most frequently classified in the OSS domain with 59, which is more than 2/3 of all papers. Examples for the OSS domain are studies using sentiment analysis on public data from platforms like GitHub, where OSS is developed (e.g. \cite{.2020}). However, there are also case studies, where the sentiments of chats from developer teams are analyzed in the industry (e.g. \cite{10.1145/3392877}) or university (e.g. \cite{Gkontzis.2017}).

~\\ \noindent
\textit{RQ 2: For what purpose is sentiment analysis used in the investigated studies?}\\
Based on their purpose, the studies on sentiment analysis in SE can be divided into three categories: Development, comparison and application. Regarding the sentiment analysis context, 52 from the $n=80$ papers focus on applying a sentiment analysis tool. In most cases, the authors of the studies wanted to perform statistical analysis to find correlations between sentiments and a specific parameters (e.g. different times of a day in bug-introducing and bug-fixing commits \cite{.2018}). Also, social aspects of developers were often studied. For example Whiting et al. \cite{10.1145/3392877} introduced a new technique that support online and remote teams in a way that the viability of the teams increase. In this context, they applied sentiment analysis on the developers' chats.

~\\ \noindent
\textit{RQ 3: What data is used as a basis for sentiment analysis?}\\
There are 48 different data sources, 6 of them are used at least three times. With 21, 20 and 20 usage respectively, JIRA, GitHub and Stack Overflow are used most frequently for training and testing sentiment analysis algorithms. Most of the data sets consist of commit/pull request comments, discussions, questions, or reviews.

When developing or comparing sentiment analysis tools, the authors of the papers often created or used data sets from platforms such as GitHub, JIRA or Stack Overflow. When using existing tools, these data sets also occurred, but most often the studies used individual data from specific development teams across all three application domains. This is also illustrated by the fact that we assigned over 3/4 of the data sets from the ``other'' category to papers of the application type. This category includes for example data sets that originate from the industrial sector (e.g. chat data from Amazon Mturk \cite{10.1145/3392877}, or student chat data from software projects in universities (e.g. \cite{Guzman.2013b,Guzman.2013}).

~\\ \noindent
\textit{RQ 4: Which approaches are used when developing sentiment analysis tools?}\\
We found 28 different existing sentiment analysis tools. SentiStrength \cite{Thelwall.2010,Thelwall.2012} stands out with 29 uses. This is probably due to the fact that the tool has been around for a while, has often been referenced as state-of-the-art, is domain-independent, and does not need to be trained as it uses a lexicon-based method.

Concerning the different machine learning methods, there are 15. SVM and Bayes stand out with 11 uses each. Neural Networks are third with 7 usages. The results indicate that SVM as well as Naive Bayes are popular machine learning methods, which were most frequently tested in the context of a sentiment analysis tool. Senti4SD \cite{Calefato.2018}, which is the most commonly used tool that is not lexicon-based, implemented an SVM because it produced the best results. Studies that compared multiple machine learning methods ended up choosing SVM most often (3). Gradient boosting ended up in second place (2).

~\\ \noindent
\textit{RQ 5: What are the difficulties of these approaches?}\\
The authors often stated that existing, domain independent tools lead to poor results in the SE domain (e.g. \cite{Calefato.2018, Imtiaz.2018, 10.1145/3180155.3180195}). This is because certain terms are used differently in the SE domain than in the non-technical context, resulting in different sentiments. In addition, the papers also described that labeling sentences with sentiment is often subjective, so people would already assign different labels to each other (e.g. \cite{9240704, Imtiaz.2018, 10.1145/3180155.3180195}).
Irony or sarcasm is also a problem that is mentioned (e.g. \cite{Islam.2017, Islam.2018b, N.Novielli.2018}), as a sentence can have a different sentiment when it is known to be meant ironically. 
Nevertheless, the listed problems should be put in a temporal context, as some of the problems have already been addressed and will be further addressed.

\subsection{Interpretation}

One possible explanation for the omnipresence of studies on OSS is that this data is available. This is supported by the fact that most of the data sets used for training or evaluation belong to the OSS domain. They are often from platforms like GitHub or Stack Overflow. This means that the public comments were mined and then manually labeled.  
But for the application of sentiment analysis in industry with the best possible performance, it would make sense to take data from the industry and train the tools with this data. Especially with the background that some papers have found that existing tools perform poorly in cross-platform settings.

One could see in the application category a tendency that there are still too few application scenarios in industry or that the tools do not yet seem interesting enough for the industrial context. Therefore, it might be necessary to investigate more intensively to what extent there is a demand for sentiment analysis in SE and to look into the reasons why it has not been used much so far in relation to OSS. One possible explanation for the scarce use in industry are legal and privacy issues. It is likely that developers have doubts if their data is analyzed according to the adequacy of the used language, and work councils often do not allow the analysis of existing data with respect to social aspects. Therefore, getting access to industry data sets is way more complicated than using data that is online available and can be used with less restrictions, if at all.

Regarding the usage of existing tools, SentiStrength \cite{Thelwall.2010,Thelwall.2012} is a well-established tool for sentiment analysis, because it has been used frequently in the application category (20) and has also constantly served as a comparison tool in the other two categories, development (1) and comparison (8). It is lexicon-based, which generally has the advantage of not having to be trained and it rarely has performance drops in different domains. The disadvantage, however, is that generic tools are not specialized for the respective domains like SE. Therefore, one of the first tools developed specifically for SE was SentiStrength-SE \cite{Islam.2017}, which was adapted to the SE domain. The advantage of the adaption to the SE domain and no need for training could explain why SentiStrength-SE \cite{Islam.2017} was used as the second most common. Other tools from the $n=80$ papers are based on traditional machine learning approaches such as SVM or Bayes, which are the most common among all papers with 11 appearances. When trained in a specific domain, Senti4SD \cite{Calefato.2018}, for example, which uses SVM, achieves better accuracy than lexicon-based tools such as SentiStrength-SE \cite{Islam.2017}, but then performs significantly worse on cross-platforms \cite{Novielli.}. However, there are also tools based on artificial neural networks, like the BERT-based model RoBERTa \cite{liu2019roberta}, which sometimes outperforms Senti4SD \cite{Calefato.2018} or other sentiment analysis tools \cite{9240704}.

Of the problems listed in Subsection \ref{sec:difficulties}, many were also attempted to be addressed by the papers, such as the adaption to the SE domain. The problem of irony or sarcasm, however, is a problem that is still not solved according to the results of our SLR. In addition, it is also mentioned that the existing tools mostly perform differently in cross-platform settings \cite{Novielli.,9240704,10.5555/2735522.2735528,Ahasanuzzaman.2020}. This means that if, for example, a tool has been trained with a GitHub data set, it will perform well on that data, but may perform poorly on data from other platforms like Stack Overflow. This should be investigated further. Possible causes could be that the choice of words is communicated differently on the platforms, e.g. different levels of politeness. Or it may also be due to the labels of the different data sets, as some papers point out that people themselves often disagree about which polarity they assign to certain sentences and the assignment is ultimately also subjective \cite{9240704,Imtiaz.2018,Kaur.2018,N.Novielli.2018,Novielli.,10.1145/3180155.3180195,Guzman.2017,Murgia.2018,Murgia.2014,Novielli.2015,S.Panichella.2015}. 
%Zitat für alle mit Irony \cite{Islam.2017,Islam.2018b,Islam.2018,N.Novielli.2018,Novielli.,Novielli.20.10.2020,ApostolosKritikos.2020,Claes.2018,Haque.2014,Souza.2017}
\subsection{Future research directions}
To address the low application of sentiment analysis in the industrial context, it can be useful to create data sets that emerge directly from industry. Based on that data, sentiment analysis tools can be created, which will potentially perform better in the field of industry, because they are be better adapted to practice. Nevertheless, there is a risk that the problems around subjective labeling and thus possibly also cross-platform performance will occur. To avoid this, it would therefore be meaningful to match the collected communication data with an additional regular sentiment survey of the developers. This way, it would be possible to compare the predictions of the trained tools with the collected manual sentiment data.

Possible solutions for the different cross-platform performance of the tools could be to examine the data sets and their labels for subjectivity, or to have them labeled according to the same emotion model like the Plutchik model \cite{PLUTCHIK19803} or the PANAS scale \cite{Watson.1988} instead of ad-hoc annotations. In addition, it might be meaningful to have it labeled by the same authors, leading to similar sets which are evaluated similarly and the bias is constantly contained in the form of a certain subjectivity, so that the tools perform more uniformly. Ad-hoc annotations may have the advantage of capturing perceived sentiments more accurately than a sentiment assignment based on emotion models. However, logically the subjectivity of an ad-hoc assignment is usually higher, so machine learning techniques then have their difficulties in achieving high performance with this subjective data. Another possibility would be to develop a tool that combines several well-performing tools to get a better result in a cross-platforming setting. Furthermore, an investigation of the expressions and politeness levels of different platforms and subdomains would be useful. It should be investigated, for example, whether the developers on GitHub communicate in a different way than on Stack Overflow. 

Logically, to address the problem of poor data, it makes sense to mine new data from the respective platforms like GitHub or Stack Overflow or even get it from the industry and label them.

The problem about irony/sarcasm handling is domain independent. Therefore, it might help to do a search of all new developments in natural language processing to see what new approaches are available to handle them.

Based on the results of our SLR, one can get good performance when using machine learning approaches like SVM or gradient boosting. 
However, our results do not contain enough data on sentiment analysis tools based on neural networks to draw conclusions about them. Nevertheless, according to the latest developments, this neural network approach delivers promising results \cite{9240704, Novielli.20.10.2020}.

\section{Conclusion}
\label{sec:conclusion}
In development teams, there is a lot communication via various channels, so the social component of a developer plays a major role. Appropriate interaction with each other in these channels is therefore of great importance. For meetings, it has been proven that a negative mood quickly affects the entire team and everyone can be demotivated afterwards. Sentiment analysis tools help to counteract this in text-based communication. 
Since, to the best of our knowledge, no SLR has been conducted to provide an overview of various tools, their development, application and problems, we have conducted such a SLR.

We analyzed 6763 papers and found 80 relevant papers. We analyzed these papers according to the application area of sentiment analysis tools, the underlying data, procedures and purpose of application. One finding is that most papers only use sentiment analysis instead of developing it or comparing its performance with other tools. 

We also identified several problems such as the handling of irony and sarcasm, the limited amount of data available to train and evaluate machine learning based tools, subjectivity in labeling data, and performance degradation in cross-platform settings. Some possible causes for the problems were mentioned as well as possible solutions. Future research should hence focus on objectively labeled data, especially on data from industry. The already existing data sets should be examined for different labels. In addition, combinations of existing tools can be considered in order to potentially achieve even better performance in a cross-platform setting.

\balance
%%
%% The next two lines define the bibliography style to be used, and
%% the bibliography file.
\bibliographystyle{ACM-Reference-Format}
\bibliography{acmart}

%%% -*-BibTeX-*-
%%% Do NOT edit. File created by BibTeX with style
%%% ACM-Reference-Format-Journals [18-Jan-2012].

\begin{thebibliography}{79}

%%% ====================================================================
%%% NOTE TO THE USER: you can override these defaults by providing
%%% customized versions of any of these macros before the \bibliography
%%% command.  Each of them MUST provide its own final punctuation,
%%% except for \shownote{}, \showDOI{}, and \showURL{}.  The latter two
%%% do not use final punctuation, in order to avoid confusing it with
%%% the Web address.
%%%
%%% To suppress output of a particular field, define its macro to expand
%%% to an empty string, or better, \unskip, like this:
%%%
%%% \newcommand{\showDOI}[1]{\unskip}   % LaTeX syntax
%%%
%%% \def \showDOI #1{\unskip}           % plain TeX syntax
%%%
%%% ====================================================================

\ifx \showCODEN    \undefined \def \showCODEN     #1{\unskip}     \fi
\ifx \showDOI      \undefined \def \showDOI       #1{#1}\fi
\ifx \showISBNx    \undefined \def \showISBNx     #1{\unskip}     \fi
\ifx \showISBNxiii \undefined \def \showISBNxiii  #1{\unskip}     \fi
\ifx \showISSN     \undefined \def \showISSN      #1{\unskip}     \fi
\ifx \showLCCN     \undefined \def \showLCCN      #1{\unskip}     \fi
\ifx \shownote     \undefined \def \shownote      #1{#1}          \fi
\ifx \showarticletitle \undefined \def \showarticletitle #1{#1}   \fi
\ifx \showURL      \undefined \def \showURL       {\relax}        \fi
% The following commands are used for tagged output and should be
% invisible to TeX
\providecommand\bibfield[2]{#2}
\providecommand\bibinfo[2]{#2}
\providecommand\natexlab[1]{#1}
\providecommand\showeprint[2][]{arXiv:#2}

\bibitem[\protect\citeauthoryear{Abo, Raj, Qazi, and Zakari}{Abo
  et~al\mbox{.}}{2019}]%
        {Mohamed.2019}
\bibfield{author}{\bibinfo{person}{Mohamed Elhag~M. Abo},
  \bibinfo{person}{Ram~Gopal Raj}, \bibinfo{person}{Atika Qazi}, {and}
  \bibinfo{person}{Abubakar Zakari}.} \bibinfo{year}{2019}\natexlab{}.
\newblock \bibinfo{title}{Sentiment Analysis for Arabic in Social Media
  Network: A Systematic Mapping Study}.
\newblock
\newblock


\bibitem[\protect\citeauthoryear{Ahasanuzzaman, Asaduzzaman, Roy, and
  Schneider}{Ahasanuzzaman et~al\mbox{.}}{2020}]%
        {Ahasanuzzaman.2020}
\bibfield{author}{\bibinfo{person}{Md Ahasanuzzaman}, \bibinfo{person}{Muhammad
  Asaduzzaman}, \bibinfo{person}{Chanchal~K. Roy}, {and}
  \bibinfo{person}{Kevin~A. Schneider}.} \bibinfo{year}{2020}\natexlab{}.
\newblock \showarticletitle{CAPS: a supervised technique for classifying Stack
  Overflow posts concerning API issues}.
\newblock \bibinfo{journal}{\emph{Empirical Software Engineering}}
  \bibinfo{volume}{25}, \bibinfo{number}{2} (\bibinfo{year}{2020}),
  \bibinfo{pages}{1493--1532}.
\newblock
\showISSN{1382-3256}
\urldef\tempurl%
\url{https://doi.org/10.1007/s10664-019-09743-4}
\showDOI{\tempurl}


\bibitem[\protect\citeauthoryear{Ahmed, Bosu, Iqbal, and Rahimi}{Ahmed
  et~al\mbox{.}}{2017}]%
        {Ahmed.2017}
\bibfield{author}{\bibinfo{person}{Toufique Ahmed}, \bibinfo{person}{Amiangshu
  Bosu}, \bibinfo{person}{Anindya Iqbal}, {and} \bibinfo{person}{Shahram
  Rahimi}.} \bibinfo{year}{2017}\natexlab{}.
\newblock \showarticletitle{SentiCR: A customized sentiment analysis tool for
  code review interactions}. In \bibinfo{booktitle}{\emph{2017 32nd IEEE/ACM
  International Conference on Automated Software Engineering (ASE)}}.
  \bibinfo{publisher}{IEEE}, \bibinfo{pages}{106--111}.
\newblock
\urldef\tempurl%
\url{https://doi.org/10.1109/ASE.2017.8115623}
\showDOI{\tempurl}


\bibitem[\protect\citeauthoryear{Aung and Myo}{Aung and Myo}{2017}]%
        {Aung.2017}
\bibfield{author}{\bibinfo{person}{Khin~Zezawar Aung} {and}
  \bibinfo{person}{Nyein~Nyein Myo}.} \bibinfo{year}{2017}\natexlab{}.
\newblock \showarticletitle{Sentiment analysis of students' comment using
  lexicon based approach}. In \bibinfo{booktitle}{\emph{16th IEEE/ACIS
  International Conference on Computer and Information Science (ICIS 2017)}},
  \bibfield{editor}{\bibinfo{person}{Guobin Zhu}} (Ed.).
  \bibinfo{publisher}{IEEE}, \bibinfo{address}{Piscataway, NJ},
  \bibinfo{pages}{149--154}.
\newblock
\showISBNx{9781509055074}
\urldef\tempurl%
\url{https://doi.org/10.1109/icis.2017.7959985}
\showDOI{\tempurl}


\bibitem[\protect\citeauthoryear{{Barbara Kitchenham} and {Stuart M.
  Charters}}{{Barbara Kitchenham} and {Stuart M. Charters}}{2007}]%
        {KitchenhamBA.2007}
\bibfield{author}{\bibinfo{person}{{Barbara Kitchenham}} {and}
  \bibinfo{person}{{Stuart M. Charters}}.} \bibinfo{year}{2007}\natexlab{}.
\newblock \bibinfo{booktitle}{\emph{Guidelines for performing Systematic
  Literature Reviews in Software Engineering}}. Vol.~\bibinfo{volume}{2}.
\newblock
\urldef\tempurl%
\url{https://www.researchgate.net/publication/302924724_Guidelines_for_performing_Systematic_Literature_Reviews_in_Software_Engineering}
\showURL{%
\tempurl}


\bibitem[\protect\citeauthoryear{Biswas, Vijay-Shanker, and Pollock}{Biswas
  et~al\mbox{.}}{2019}]%
        {Biswas.2019}
\bibfield{author}{\bibinfo{person}{Eeshita Biswas}, \bibinfo{person}{K.
  Vijay-Shanker}, {and} \bibinfo{person}{Lori Pollock}.}
  \bibinfo{year}{2019}\natexlab{}.
\newblock \showarticletitle{Exploring Word Embedding Techniques to Improve
  Sentiment Analysis of Software Engineering Texts}. In
  \bibinfo{booktitle}{\emph{2019 IEEE/ACM 16th International Conference on
  Mining Software Repositories}}. \bibinfo{publisher}{IEEE},
  \bibinfo{address}{Piscataway, NJ}.
\newblock
\showISBNx{9781728134123}
\urldef\tempurl%
\url{https://doi.org/10.1109/msr.2019.00020}
\showDOI{\tempurl}


\bibitem[\protect\citeauthoryear{Cagnoni, Cozzini, Lombardo, Mordonini, Poggi,
  and Tomaiuolo}{Cagnoni et~al\mbox{.}}{2020}]%
        {Cagnoni.2020}
\bibfield{author}{\bibinfo{person}{Stefano Cagnoni}, \bibinfo{person}{Lorenzo
  Cozzini}, \bibinfo{person}{Gianfranco Lombardo}, \bibinfo{person}{Monica
  Mordonini}, \bibinfo{person}{Agostino Poggi}, {and} \bibinfo{person}{Michele
  Tomaiuolo}.} \bibinfo{year}{2020}\natexlab{}.
\newblock \showarticletitle{Emotion-based analysis of programming languages on
  Stack Overflow}.
\newblock \bibinfo{journal}{\emph{ICT Express}} \bibinfo{volume}{6},
  \bibinfo{number}{3} (\bibinfo{year}{2020}), \bibinfo{pages}{238--242}.
\newblock
\showISSN{2405-9595}
\urldef\tempurl%
\url{https://doi.org/10.1016/j.icte.2020.07.002}
\showDOI{\tempurl}


\bibitem[\protect\citeauthoryear{Calefato, Lanubile, Maiorano, and
  Novielli}{Calefato et~al\mbox{.}}{2018}]%
        {Calefato.2018}
\bibfield{author}{\bibinfo{person}{Fabio Calefato}, \bibinfo{person}{Filippo
  Lanubile}, \bibinfo{person}{Federico Maiorano}, {and} \bibinfo{person}{Nicole
  Novielli}.} \bibinfo{year}{2018}\natexlab{}.
\newblock \showarticletitle{Sentiment Polarity Detection for Software
  Development}.
\newblock \bibinfo{journal}{\emph{Empirical Software Engineering}}
  \bibinfo{volume}{23}, \bibinfo{number}{3} (\bibinfo{year}{2018}),
  \bibinfo{pages}{1352--1382}.
\newblock
\showISSN{1382-3256}
\urldef\tempurl%
\url{https://doi.org/10.1007/s10664-017-9546-9}
\showDOI{\tempurl}


\bibitem[\protect\citeauthoryear{Calefato, Lanubile, and Novielli}{Calefato
  et~al\mbox{.}}{2017}]%
        {Calefato.2017}
\bibfield{author}{\bibinfo{person}{Fabio Calefato}, \bibinfo{person}{Filippo
  Lanubile}, {and} \bibinfo{person}{Nicole Novielli}.}
  \bibinfo{year}{2017}\natexlab{}.
\newblock \showarticletitle{EmoTxt: A toolkit for emotion recognition from
  text}. In \bibinfo{booktitle}{\emph{2017 Seventh International Conference on
  Affective Computing and Intelligent Interaction Workshops and Demos
  (ACIIW)}}. \bibinfo{publisher}{IEEE}, \bibinfo{address}{Piscataway, NJ},
  \bibinfo{pages}{79--80}.
\newblock
\showISBNx{9781538606803}
\urldef\tempurl%
\url{https://doi.org/10.1109/ACIIW.2017.8272591}
\showDOI{\tempurl}


\bibitem[\protect\citeauthoryear{Chen, Cao, Lu, Mei, and Liu}{Chen
  et~al\mbox{.}}{2019}]%
        {Chen.2019}
\bibfield{author}{\bibinfo{person}{Zhenpeng Chen}, \bibinfo{person}{Yanbin
  Cao}, \bibinfo{person}{Xuan Lu}, \bibinfo{person}{Qiaozhu Mei}, {and}
  \bibinfo{person}{Xuanzhe Liu}.} \bibinfo{year}{2019}\natexlab{}.
\newblock \showarticletitle{SEntiMoji: An Emoji-Powered Learning Approach for
  Sentiment Analysis in Software Engineering}. In
  \bibinfo{booktitle}{\emph{Proceedings of the 2019 27th ACM Joint Meeting on
  European Software Engineering Conference and Symposium on the Foundations of
  Software Engineering}} (Tallinn, Estonia) \emph{(\bibinfo{series}{ESEC/FSE
  2019})}. \bibinfo{publisher}{Association for Computing Machinery},
  \bibinfo{address}{New York, NY, USA}, \bibinfo{pages}{841–852}.
\newblock
\showISBNx{9781450355728}
\urldef\tempurl%
\url{https://doi.org/10.1145/3338906.3338977}
\showDOI{\tempurl}


\bibitem[\protect\citeauthoryear{De~Choudhury and Counts}{De~Choudhury and
  Counts}{2013}]%
        {10.1145/2441776.2441812}
\bibfield{author}{\bibinfo{person}{Munmun De~Choudhury} {and}
  \bibinfo{person}{Scott Counts}.} \bibinfo{year}{2013}\natexlab{}.
\newblock \bibinfo{booktitle}{\emph{Understanding Affect in the Workplace via
  Social Media}}.
\newblock \bibinfo{publisher}{Association for Computing Machinery},
  \bibinfo{address}{New York, NY, USA}, \bibinfo{pages}{303–316}.
\newblock
\showISBNx{9781450313315}
\urldef\tempurl%
\url{https://doi.org/10.1145/2441776.2441812}
\showURL{%
\tempurl}


\bibitem[\protect\citeauthoryear{Deshpande and Rao}{Deshpande and Rao}{2017}]%
        {8389299}
\bibfield{author}{\bibinfo{person}{Mandar Deshpande} {and}
  \bibinfo{person}{Vignesh Rao}.} \bibinfo{year}{2017}\natexlab{}.
\newblock \showarticletitle{Depression detection using emotion artificial
  intelligence}. In \bibinfo{booktitle}{\emph{2017 International Conference on
  Intelligent Sustainable Systems (ICISS)}}. \bibinfo{pages}{858--862}.
\newblock
\urldef\tempurl%
\url{https://doi.org/10.1109/ISS1.2017.8389299}
\showDOI{\tempurl}


\bibitem[\protect\citeauthoryear{Devika, Sunitha, and Ganesh}{Devika
  et~al\mbox{.}}{2016}]%
        {Devika.2016}
\bibfield{author}{\bibinfo{person}{M.D. Devika}, \bibinfo{person}{C. Sunitha},
  {and} \bibinfo{person}{Amal Ganesh}.} \bibinfo{year}{2016}\natexlab{}.
\newblock \showarticletitle{Sentiment Analysis: A Comparative Study on
  Different Approaches}.
\newblock \bibinfo{journal}{\emph{Procedia Computer Science}}
  \bibinfo{volume}{87} (\bibinfo{year}{2016}), \bibinfo{pages}{44 -- 49}.
\newblock
\showISSN{1877-0509}
\urldef\tempurl%
\url{https://doi.org/10.1016/j.procs.2016.05.124}
\showDOI{\tempurl}
\newblock
\shownote{Fourth International Conference on Recent Trends in Computer Science
  \& Engineering (ICRTCSE 2016).}


\bibitem[\protect\citeauthoryear{Devlin, Chang, Lee, and Toutanova}{Devlin
  et~al\mbox{.}}{2019}]%
        {Devlin.2019}
\bibfield{author}{\bibinfo{person}{Jacob Devlin}, \bibinfo{person}{Ming-Wei
  Chang}, \bibinfo{person}{Kenton Lee}, {and} \bibinfo{person}{Kristina
  Toutanova}.} \bibinfo{year}{2019}\natexlab{}.
\newblock \bibinfo{title}{BERT: Pre-training of Deep Bidirectional Transformers
  for Language Understanding}.
\newblock
\newblock


\bibitem[\protect\citeauthoryear{Ding, Sun, Wang, and Liu}{Ding
  et~al\mbox{.}}{2018}]%
        {Ding.2018}
\bibfield{author}{\bibinfo{person}{Jin Ding}, \bibinfo{person}{Hailong Sun},
  \bibinfo{person}{Xu Wang}, {and} \bibinfo{person}{Xudong Liu}.}
  \bibinfo{year}{2018}\natexlab{}.
\newblock \showarticletitle{Entity-level sentiment analysis of issue comments}.
  In \bibinfo{booktitle}{\emph{2018 ACM/IEEE 3rd International Workshop on
  Emotion Awareness in Software Engineering}}. \bibinfo{publisher}{IEEE},
  \bibinfo{address}{Piscataway, NJ}.
\newblock
\showISBNx{9781450357517}
\urldef\tempurl%
\url{https://doi.org/10.1145/3194932.3194935}
\showDOI{\tempurl}


\bibitem[\protect\citeauthoryear{El-Halees}{El-Halees}{2014}]%
        {ElHalees.2014}
\bibfield{author}{\bibinfo{person}{Alaa~Mustafa El-Halees}.}
  \bibinfo{year}{2014}\natexlab{}.
\newblock \showarticletitle{Software Usability Evaluation Using Opinion
  Mining}.
\newblock \bibinfo{journal}{\emph{Journal of Software}} \bibinfo{volume}{9},
  \bibinfo{number}{2} (\bibinfo{year}{2014}).
\newblock
\showISSN{1796-217X}
\urldef\tempurl%
\url{https://doi.org/10.4304/jsw.9.2.343-349}
\showDOI{\tempurl}


\bibitem[\protect\citeauthoryear{Fang and Zhan}{Fang and Zhan}{2015}]%
        {Fang.2015}
\bibfield{author}{\bibinfo{person}{Xing Fang} {and} \bibinfo{person}{Justin
  Zhan}.} \bibinfo{year}{2015}\natexlab{}.
\newblock \showarticletitle{Sentiment analysis using product review data}.
\newblock \bibinfo{journal}{\emph{Journal of Big Data}} \bibinfo{volume}{2},
  \bibinfo{number}{1} (\bibinfo{year}{2015}), \bibinfo{pages}{1--14}.
\newblock
\showISSN{2196-1115}
\urldef\tempurl%
\url{https://doi.org/10.1186/s40537-015-0015-2}
\showDOI{\tempurl}


\bibitem[\protect\citeauthoryear{Garousi, Petersen, and Ozkan}{Garousi
  et~al\mbox{.}}{2016}]%
        {GAROUSI2016106}
\bibfield{author}{\bibinfo{person}{Vahid Garousi}, \bibinfo{person}{Kai
  Petersen}, {and} \bibinfo{person}{Baris Ozkan}.}
  \bibinfo{year}{2016}\natexlab{}.
\newblock \showarticletitle{Challenges and best practices in industry-academia
  collaborations in software engineering: A systematic literature review}.
\newblock \bibinfo{journal}{\emph{Information and Software Technology}}
  \bibinfo{volume}{79} (\bibinfo{year}{2016}), \bibinfo{pages}{106--127}.
\newblock
\showISSN{0950-5849}
\urldef\tempurl%
\url{https://doi.org/10.1016/j.infsof.2016.07.006}
\showDOI{\tempurl}


\bibitem[\protect\citeauthoryear{Gkontzis, Karachristos, Panagiotakopoulos,
  Stavropoulos, and Verykios}{Gkontzis et~al\mbox{.}}{2017}]%
        {Gkontzis.2017}
\bibfield{author}{\bibinfo{person}{Andreas~F. Gkontzis},
  \bibinfo{person}{Christoforos~V. Karachristos}, \bibinfo{person}{Chris~T.
  Panagiotakopoulos}, \bibinfo{person}{Elias~C. Stavropoulos}, {and}
  \bibinfo{person}{Vassilios~S. Verykios}.} \bibinfo{year}{2017}\natexlab{}.
\newblock \showarticletitle{Sentiment Analysis to Track Emotion and Polarity in
  Student Fora}. In \bibinfo{booktitle}{\emph{Proceedings of the 21st
  Pan-Hellenic Conference on Informatics}},
  \bibfield{editor}{\bibinfo{person}{Vasileios Vlachos}} (Ed.).
  \bibinfo{publisher}{ACM}, \bibinfo{address}{New York, NY}.
\newblock
\showISBNx{9781450353557}
\urldef\tempurl%
\url{https://doi.org/10.1145/3139367.3139389}
\showDOI{\tempurl}


\bibitem[\protect\citeauthoryear{Graziotin, Wang, and Abrahamsson}{Graziotin
  et~al\mbox{.}}{2014}]%
        {Graziotin.2014}
\bibfield{author}{\bibinfo{person}{Daniel Graziotin}, \bibinfo{person}{Xiaofeng
  Wang}, {and} \bibinfo{person}{Pekka Abrahamsson}.}
  \bibinfo{year}{2014}\natexlab{}.
\newblock \showarticletitle{Happy software developers solve problems better:
  psychological measurements in empirical software engineering}.
\newblock \bibinfo{journal}{\emph{PeerJ}}  \bibinfo{volume}{2}
  (\bibinfo{year}{2014}), \bibinfo{pages}{e289}.
\newblock
\showISSN{2167-8359}
\urldef\tempurl%
\url{https://doi.org/10.7717/peerj.289}
\showDOI{\tempurl}


\bibitem[\protect\citeauthoryear{Graziotin, Wang, and Abrahamsson}{Graziotin
  et~al\mbox{.}}{2015}]%
        {10.1002/smr.1673}
\bibfield{author}{\bibinfo{person}{Daniel Graziotin}, \bibinfo{person}{Xiaofeng
  Wang}, {and} \bibinfo{person}{Pekka Abrahamsson}.}
  \bibinfo{year}{2015}\natexlab{}.
\newblock \showarticletitle{Do feelings matter? On the correlation of affects
  and the self-assessed productivity in software engineering}.
\newblock \bibinfo{journal}{\emph{Journal of Software: Evolution and Process}}
  \bibinfo{volume}{27}, \bibinfo{number}{7} (\bibinfo{year}{2015}),
  \bibinfo{pages}{467--487}.
\newblock
\urldef\tempurl%
\url{https://doi.org/10.1002/smr.1673}
\showDOI{\tempurl}


\bibitem[\protect\citeauthoryear{Guzman}{Guzman}{2013}]%
        {Guzman.2013b}
\bibfield{author}{\bibinfo{person}{Emitza Guzman}.}
  \bibinfo{year}{2013}\natexlab{}.
\newblock \showarticletitle{Visualizing emotions in software development
  projects}. In \bibinfo{booktitle}{\emph{2013 First IEEE Working Conference on
  Software Visualization (VISSOFT)}},
  \bibfield{editor}{\bibinfo{person}{Alexandru Telea}} (Ed.).
  \bibinfo{publisher}{IEEE}, \bibinfo{address}{Piscataway, NJ}.
\newblock
\showISBNx{9781479914579}
\urldef\tempurl%
\url{https://doi.org/10.1109/vissoft.2013.6650529}
\showDOI{\tempurl}


\bibitem[\protect\citeauthoryear{Guzman, Alkadhi, and Seyff}{Guzman
  et~al\mbox{.}}{2017}]%
        {Guzman.2017}
\bibfield{author}{\bibinfo{person}{Emitza Guzman}, \bibinfo{person}{Rana
  Alkadhi}, {and} \bibinfo{person}{Norbert Seyff}.}
  \bibinfo{year}{2017}\natexlab{}.
\newblock \showarticletitle{An exploratory study of Twitter messages about
  software applications}.
\newblock \bibinfo{journal}{\emph{Requirements Engineering}}
  \bibinfo{volume}{22}, \bibinfo{number}{3} (\bibinfo{year}{2017}),
  \bibinfo{pages}{387--412}.
\newblock
\showISSN{1432-010X}
\urldef\tempurl%
\url{https://doi.org/10.1007/s00766-017-0274-x}
\showDOI{\tempurl}


\bibitem[\protect\citeauthoryear{Guzman, Az{\'o}car, and Li}{Guzman
  et~al\mbox{.}}{2014}]%
        {Guzman.2014}
\bibfield{author}{\bibinfo{person}{Emitza Guzman}, \bibinfo{person}{David
  Az{\'o}car}, {and} \bibinfo{person}{Yang Li}.}
  \bibinfo{year}{2014}\natexlab{}.
\newblock \showarticletitle{Sentiment analysis of commit comments in GitHub: an
  empirical study}. In \bibinfo{booktitle}{\emph{11th Working Conference on
  Mining Software Repositories : proceedings : May 31 - June 1, 2014,
  Hyderabad, India}}, \bibfield{editor}{\bibinfo{person}{Sung Kim},
  \bibinfo{person}{Martin Pinzger}, {and} \bibinfo{person}{Premkumar Devanbu}}
  (Eds.). \bibinfo{publisher}{ACM}, \bibinfo{address}{[Place of publication not
  identified]}, \bibinfo{pages}{352--355}.
\newblock
\showISBNx{9781450328630}
\urldef\tempurl%
\url{https://doi.org/10.1145/2597073.2597118}
\showDOI{\tempurl}


\bibitem[\protect\citeauthoryear{Guzman and Bruegge}{Guzman and
  Bruegge}{2013}]%
        {Guzman.2013}
\bibfield{author}{\bibinfo{person}{Emitza Guzman} {and} \bibinfo{person}{Bernd
  Bruegge}.} \bibinfo{year}{2013}\natexlab{}.
\newblock \showarticletitle{Towards Emotional Awareness in Software Development
  Teams}. In \bibinfo{booktitle}{\emph{2013 9th Joint Meeting of the European
  Software Engineering Conference and the ACM SIGSOFT Symposium on the
  Foundations of Software Engineering (ESEC/FSE) : proceedings : August 18-26,
  2013, Saint Petersburg, Russia}} (Saint Petersburg, Russia)
  \emph{(\bibinfo{series}{ESEC/FSE 2013})},
  \bibfield{editor}{\bibinfo{person}{Bertrand Meyer}, \bibinfo{person}{Mira
  Mezini}, {and} \bibinfo{person}{Luciano Baresi}} (Eds.).
  \bibinfo{publisher}{ACM}, \bibinfo{pages}{671–674}.
\newblock
\showISBNx{9781450322379}
\urldef\tempurl%
\url{https://doi.org/10.1145/2491411.2494578}
\showDOI{\tempurl}


\bibitem[\protect\citeauthoryear{Holmes, Donkin, and Witten}{Holmes
  et~al\mbox{.}}{1994}]%
        {396988}
\bibfield{author}{\bibinfo{person}{G. Holmes}, \bibinfo{person}{A. Donkin},
  {and} \bibinfo{person}{I.H. Witten}.} \bibinfo{year}{1994}\natexlab{}.
\newblock \showarticletitle{WEKA: a machine learning workbench}. In
  \bibinfo{booktitle}{\emph{Proceedings of ANZIIS '94 - Australian New Zealnd
  Intelligent Information Systems Conference}}. \bibinfo{pages}{357--361}.
\newblock
\urldef\tempurl%
\url{https://doi.org/10.1109/ANZIIS.1994.396988}
\showDOI{\tempurl}


\bibitem[\protect\citeauthoryear{Imtiaz, Middleton, Girouard, and
  Murphy-Hill}{Imtiaz et~al\mbox{.}}{2018}]%
        {Imtiaz.2018}
\bibfield{author}{\bibinfo{person}{Nasif Imtiaz}, \bibinfo{person}{Justin
  Middleton}, \bibinfo{person}{Peter Girouard}, {and} \bibinfo{person}{Emerson
  Murphy-Hill}.} \bibinfo{year}{2018}\natexlab{}.
\newblock \showarticletitle{Sentiment and politeness analysis tools on
  developer discussions are unreliable, but so are people}. In
  \bibinfo{booktitle}{\emph{2018 ACM/IEEE 3rd International Workshop on Emotion
  Awareness in Software Engineering}}. \bibinfo{publisher}{IEEE},
  \bibinfo{address}{Piscataway, NJ}.
\newblock
\showISBNx{9781450357517}
\urldef\tempurl%
\url{https://doi.org/10.1145/3194932.3194938}
\showDOI{\tempurl}


\bibitem[\protect\citeauthoryear{Islam and Zibran}{Islam and Zibran}{2017}]%
        {Islam.2017}
\bibfield{author}{\bibinfo{person}{Md~Rakibul Islam} {and}
  \bibinfo{person}{Minhaz~F. Zibran}.} \bibinfo{year}{2017}\natexlab{}.
\newblock \showarticletitle{Leveraging Automated Sentiment Analysis in Software
  Engineering}. In \bibinfo{booktitle}{\emph{2017 IEEE/ACM 14th International
  Conference on Mining Software Repositories}}. \bibinfo{publisher}{IEEE},
  \bibinfo{address}{Piscataway, NJ}.
\newblock
\showISBNx{9781538615447}
\urldef\tempurl%
\url{https://doi.org/10.1109/msr.2017.9}
\showDOI{\tempurl}


\bibitem[\protect\citeauthoryear{Islam and Zibran}{Islam and Zibran}{2018a}]%
        {Islam.2018c}
\bibfield{author}{\bibinfo{person}{Md~Rakibul Islam} {and}
  \bibinfo{person}{Minhaz~F. Zibran}.} \bibinfo{year}{2018}\natexlab{a}.
\newblock \showarticletitle{A comparison of software engineering domain
  specific sentiment analysis tools}. In \bibinfo{booktitle}{\emph{25th IEEE
  International Conference on Software Analysis, Evolution and Reengineering}}.
  \bibinfo{publisher}{IEEE}, \bibinfo{address}{Piscataway, NJ}.
\newblock
\showISBNx{9781538649695}
\urldef\tempurl%
\url{https://doi.org/10.1109/saner.2018.8330245}
\showDOI{\tempurl}


\bibitem[\protect\citeauthoryear{Islam and Zibran}{Islam and Zibran}{2018b}]%
        {Islam.2018b}
\bibfield{author}{\bibinfo{person}{Md~Rakibul Islam} {and}
  \bibinfo{person}{Minhaz~F. Zibran}.} \bibinfo{year}{2018}\natexlab{b}.
\newblock \showarticletitle{DEVA: sensing emotions in the valence arousal space
  in software engineering text}. In \bibinfo{booktitle}{\emph{Applied computing
  2018}}, \bibfield{editor}{\bibinfo{person}{Hisham~M. Haddad},
  \bibinfo{person}{Roger~L. Wainwright}, {and} \bibinfo{person}{Richard
  Chbeir}} (Eds.). \bibinfo{publisher}{{Association for Computing Machinery
  Inc. (ACM)}}, \bibinfo{address}{New York, NY}, \bibinfo{pages}{1536--1543}.
\newblock
\showISBNx{9781450351911}
\urldef\tempurl%
\url{https://doi.org/10.1145/3167132.3167296}
\showDOI{\tempurl}


\bibitem[\protect\citeauthoryear{Islam and Zibran}{Islam and Zibran}{2018c}]%
        {.2018}
\bibfield{author}{\bibinfo{person}{Md~Rabikul Islam} {and}
  \bibinfo{person}{Minhaz~F. Zibran}.} \bibinfo{year}{2018}\natexlab{c}.
\newblock \bibinfo{booktitle}{\emph{Sentiment Analysis of Software Bug Related
  Commit Messages}}.
\newblock \bibinfo{publisher}{ISCA}. 3--8 pages.
\newblock
\showISBNx{978-1-943436-05-7}
\urldef\tempurl%
\url{https://www.searchdl.org/resources/public/conf/2018/sede/10417.pdf}
\showURL{%
\tempurl}


\bibitem[\protect\citeauthoryear{Islam and Zibran}{Islam and Zibran}{2018d}]%
        {Islam.2018}
\bibfield{author}{\bibinfo{person}{Md~Rakibul Islam} {and}
  \bibinfo{person}{Minhaz~F. Zibran}.} \bibinfo{year}{2018}\natexlab{d}.
\newblock \showarticletitle{SentiStrength-SE: Exploiting domain specificity for
  improved sentiment analysis in software engineering text}.
\newblock \bibinfo{journal}{\emph{Journal of Systems and Software}}
  \bibinfo{volume}{145} (\bibinfo{year}{2018}), \bibinfo{pages}{125--146}.
\newblock
\showISSN{01641212}
\urldef\tempurl%
\url{https://doi.org/10.1016/j.jss.2018.08.030}
\showDOI{\tempurl}


\bibitem[\protect\citeauthoryear{Kaewyong, Sukprasert, Salim, and
  Phang}{Kaewyong et~al\mbox{.}}{2015}]%
        {.2015}
\bibfield{author}{\bibinfo{person}{Phuripoj Kaewyong}, \bibinfo{person}{Anupong
  Sukprasert}, \bibinfo{person}{Naomie Salim}, {and} \bibinfo{person}{Fatin
  Phang}.} \bibinfo{year}{2015}\natexlab{}.
\newblock \showarticletitle{The possibility of students' comments automatic
  interpret using lexicon based sentiment analysis to teacher evaluation}.
\newblock
\urldef\tempurl%
\url{https://www.researchgate.net/publication/285581082_THE_POSSIBILITY_OF_STUDENTS\%27_COMMENTS_AUTOMATIC_INTERPRET_USING_LEXICON_BASED_SENTIMENT_ANALYSIS_TO_TEACHER_EVALUATION}
\showURL{%
\tempurl}


\bibitem[\protect\citeauthoryear{Kaur, Singh, Dhillon, and Bisht}{Kaur
  et~al\mbox{.}}{2018}]%
        {Kaur.2018}
\bibfield{author}{\bibinfo{person}{Arvinder Kaur}, \bibinfo{person}{Amrit~Pal
  Singh}, \bibinfo{person}{Guneet~Singh Dhillon}, {and} \bibinfo{person}{Divesh
  Bisht}.} \bibinfo{year}{2018}\natexlab{}.
\newblock \showarticletitle{Emotion Mining and Sentiment Analysis in Software
  Engineering Domain}. In \bibinfo{booktitle}{\emph{Proceedings of the Second
  International Conference on Electronics, Communication and Aerospace
  Technology (ICECA 2018)}}. \bibinfo{publisher}{IEEE},
  \bibinfo{address}{Piscataway, NJ}, \bibinfo{pages}{1170--1173}.
\newblock
\showISBNx{978-1-5386-0965-1}
\urldef\tempurl%
\url{https://doi.org/10.1109/ICECA.2018.8474619}
\showDOI{\tempurl}


\bibitem[\protect\citeauthoryear{Kitchenham, {Pearl Brereton}, Budgen, Turner,
  Bailey, and Linkman}{Kitchenham et~al\mbox{.}}{2009}]%
        {KITCHENHAM20097}
\bibfield{author}{\bibinfo{person}{Barbara Kitchenham}, \bibinfo{person}{O.
  {Pearl Brereton}}, \bibinfo{person}{David Budgen}, \bibinfo{person}{Mark
  Turner}, \bibinfo{person}{John Bailey}, {and} \bibinfo{person}{Stephen
  Linkman}.} \bibinfo{year}{2009}\natexlab{}.
\newblock \showarticletitle{Systematic literature reviews in software
  engineering – A systematic literature review}.
\newblock \bibinfo{journal}{\emph{Information and Software Technology}}
  \bibinfo{volume}{51}, \bibinfo{number}{1} (\bibinfo{year}{2009}),
  \bibinfo{pages}{7 -- 15}.
\newblock
\showISSN{0950-5849}
\urldef\tempurl%
\url{https://doi.org/10.1016/j.infsof.2008.09.009}
\showDOI{\tempurl}
\newblock
\shownote{Special Section - Most Cited Articles in 2002 and Regular Research
  Papers.}


\bibitem[\protect\citeauthoryear{Kl{\"u}nder, Hohl, Prenner, and
  Schneider}{Kl{\"u}nder et~al\mbox{.}}{2019}]%
        {Klunder.2019}
\bibfield{author}{\bibinfo{person}{Jil Ann-Christin Kl{\"u}nder},
  \bibinfo{person}{Philipp Hohl}, \bibinfo{person}{Nils Prenner}, {and}
  \bibinfo{person}{Kurt Schneider}.} \bibinfo{year}{2019}\natexlab{}.
\newblock \showarticletitle{Transformation towards agile software product line
  engineering in large companies: A literature review}.
\newblock \bibinfo{journal}{\emph{Journal of Software: Evolution and Process}}
  \bibinfo{volume}{31}, \bibinfo{number}{5} (\bibinfo{year}{2019}),
  \bibinfo{pages}{e2168}.
\newblock
\showISSN{2047-7481}
\urldef\tempurl%
\url{https://doi.org/10.1002/smr.2168}
\showDOI{\tempurl}


\bibitem[\protect\citeauthoryear{Kosa, Yilmaz, O’Connor, and Clarke}{Kosa
  et~al\mbox{.}}{2016}]%
        {kosa.2016}
\bibfield{author}{\bibinfo{person}{Mehmet Kosa}, \bibinfo{person}{Murat
  Yilmaz}, \bibinfo{person}{Rory O’Connor}, {and} \bibinfo{person}{Paul
  Clarke}.} \bibinfo{year}{2016}\natexlab{}.
\newblock \showarticletitle{Software Engineering Education and Games: A
  Systematic Literature Review}.
\newblock \bibinfo{journal}{\emph{Journal of Universal Computer Science}}
  \bibinfo{volume}{22} (\bibinfo{date}{12} \bibinfo{year}{2016}),
  \bibinfo{pages}{1558–1574}.
\newblock


\bibitem[\protect\citeauthoryear{Kraut and Streeter}{Kraut and
  Streeter}{1995}]%
        {10.1145/203330.203345}
\bibfield{author}{\bibinfo{person}{Robert~E. Kraut} {and}
  \bibinfo{person}{Lynn~A. Streeter}.} \bibinfo{year}{1995}\natexlab{}.
\newblock \showarticletitle{Coordination in Software Development}.
\newblock \bibinfo{journal}{\emph{Commun. ACM}} \bibinfo{volume}{38},
  \bibinfo{number}{3} (\bibinfo{date}{March} \bibinfo{year}{1995}),
  \bibinfo{pages}{69–81}.
\newblock
\showISSN{0001-0782}
\urldef\tempurl%
\url{https://doi.org/10.1145/203330.203345}
\showDOI{\tempurl}


\bibitem[\protect\citeauthoryear{Kuhrmann, Tell, Kl{\"u}nder, Hebig, {Sherlock
  A. Licorish}, and MacDonell}{Kuhrmann et~al\mbox{.}}{2018}]%
        {Kuhrmann.2018}
\bibfield{author}{\bibinfo{person}{Marco Kuhrmann}, \bibinfo{person}{Paolo
  Tell}, \bibinfo{person}{Jil Kl{\"u}nder}, \bibinfo{person}{Regina Hebig},
  \bibinfo{person}{{Sherlock A. Licorish}}, {and} \bibinfo{person}{Stephen~G.
  MacDonell}.} \bibinfo{year}{2018}\natexlab{}.
\newblock \bibinfo{title}{HELENA Stage 2 Results}.
\newblock
\newblock
\urldef\tempurl%
\url{https://doi.org/10.13140/RG.2.2.14807.52649}
\showDOI{\tempurl}


\bibitem[\protect\citeauthoryear{Kumar and Jaiswal}{Kumar and Jaiswal}{2020}]%
        {Kumar.2020}
\bibfield{author}{\bibinfo{person}{Akshi Kumar} {and} \bibinfo{person}{Arunima
  Jaiswal}.} \bibinfo{year}{2020}\natexlab{}.
\newblock \showarticletitle{Systematic literature review of sentiment analysis
  on Twitter using soft computing techniques}.
\newblock \bibinfo{journal}{\emph{Concurrency and Computation: Practice and
  Experience}} \bibinfo{volume}{32}, \bibinfo{number}{1}
  (\bibinfo{year}{2020}), \bibinfo{pages}{e5107}.
\newblock
\urldef\tempurl%
\url{https://doi.org/10.1002/cpe.5107}
\showDOI{\tempurl}
\newblock
\shownote{e5107 CPE-18-1167.R1.}


\bibitem[\protect\citeauthoryear{Lin, Zampetti, Bavota, {Di Penta}, and
  Lanza}{Lin et~al\mbox{.}}{2019}]%
        {Lin.2019}
\bibfield{author}{\bibinfo{person}{Bin Lin}, \bibinfo{person}{Fiorella
  Zampetti}, \bibinfo{person}{Gabriele Bavota}, \bibinfo{person}{Massimiliano
  {Di Penta}}, {and} \bibinfo{person}{Michele Lanza}.}
  \bibinfo{year}{2019}\natexlab{}.
\newblock \showarticletitle{Pattern-Based Mining of Opinions in Q{\&}A
  Websites}. In \bibinfo{booktitle}{\emph{2019 IEEE/ACM 41st International
  Conference on Software Engineering}}. \bibinfo{publisher}{IEEE},
  \bibinfo{address}{Piscataway, NJ}.
\newblock
\showISBNx{9781728108698}
\urldef\tempurl%
\url{https://doi.org/10.1109/icse.2019.00066}
\showDOI{\tempurl}


\bibitem[\protect\citeauthoryear{Lin, Zampetti, Bavota, Di~Penta, Lanza, and
  Oliveto}{Lin et~al\mbox{.}}{2018}]%
        {10.1145/3180155.3180195}
\bibfield{author}{\bibinfo{person}{Bin Lin}, \bibinfo{person}{Fiorella
  Zampetti}, \bibinfo{person}{Gabriele Bavota}, \bibinfo{person}{Massimiliano
  Di~Penta}, \bibinfo{person}{Michele Lanza}, {and} \bibinfo{person}{Rocco
  Oliveto}.} \bibinfo{year}{2018}\natexlab{}.
\newblock \showarticletitle{Sentiment Analysis for Software Engineering: How
  Far Can We Go?}. In \bibinfo{booktitle}{\emph{Proceedings of the 40th
  International Conference on Software Engineering}} (Gothenburg, Sweden)
  \emph{(\bibinfo{series}{ICSE '18})}. \bibinfo{publisher}{Association for
  Computing Machinery}, \bibinfo{address}{New York, NY, USA},
  \bibinfo{pages}{94–104}.
\newblock
\showISBNx{9781450356381}
\urldef\tempurl%
\url{https://doi.org/10.1145/3180155.3180195}
\showDOI{\tempurl}


\bibitem[\protect\citeauthoryear{Liu}{Liu}{2012}]%
        {Liu.2012}
\bibfield{author}{\bibinfo{person}{Bing Liu}.} \bibinfo{year}{2012}\natexlab{}.
\newblock \showarticletitle{Sentiment Analysis and Opinion Mining}.
\newblock \bibinfo{journal}{\emph{Synthesis Lectures on Human Language
  Technologies}} \bibinfo{volume}{5}, \bibinfo{number}{1}
  (\bibinfo{year}{2012}), \bibinfo{pages}{1--167}.
\newblock
\showISSN{1947-4040}
\urldef\tempurl%
\url{https://doi.org/10.2200/S00416ED1V01Y201204HLT016}
\showDOI{\tempurl}


\bibitem[\protect\citeauthoryear{Liu and Zhang}{Liu and Zhang}{2012}]%
        {Liu2012}
\bibfield{author}{\bibinfo{person}{Bing Liu} {and} \bibinfo{person}{Lei
  Zhang}.} \bibinfo{year}{2012}\natexlab{}.
\newblock \bibinfo{booktitle}{\emph{A Survey of Opinion Mining and Sentiment
  Analysis}}.
\newblock \bibinfo{publisher}{Springer US}, \bibinfo{address}{Boston, MA},
  \bibinfo{pages}{415--463}.
\newblock
\showISBNx{978-1-4614-3223-4}
\urldef\tempurl%
\url{https://doi.org/10.1007/978-1-4614-3223-4_13}
\showDOI{\tempurl}


\bibitem[\protect\citeauthoryear{Liu, Ott, Goyal, Du, Joshi, Chen, Levy, Lewis,
  Zettlemoyer, and Stoyanov}{Liu et~al\mbox{.}}{2019}]%
        {liu2019roberta}
\bibfield{author}{\bibinfo{person}{Yinhan Liu}, \bibinfo{person}{Myle Ott},
  \bibinfo{person}{Naman Goyal}, \bibinfo{person}{Jingfei Du},
  \bibinfo{person}{Mandar Joshi}, \bibinfo{person}{Danqi Chen},
  \bibinfo{person}{Omer Levy}, \bibinfo{person}{Mike Lewis},
  \bibinfo{person}{Luke Zettlemoyer}, {and} \bibinfo{person}{Veselin
  Stoyanov}.} \bibinfo{year}{2019}\natexlab{}.
\newblock \bibinfo{title}{RoBERTa: A Robustly Optimized BERT Pretraining
  Approach}.
\newblock
\newblock


\bibitem[\protect\citeauthoryear{Loper and Bird}{Loper and Bird}{2002}]%
        {Loper.17.05.2002}
\bibfield{author}{\bibinfo{person}{Edward Loper} {and} \bibinfo{person}{Steven
  Bird}.} \bibinfo{year}{2002}\natexlab{}.
\newblock \bibinfo{title}{NLTK: The Natural Language Toolkit}.
\newblock
\newblock
\urldef\tempurl%
\url{https://arxiv.org/pdf/cs/0205028}
\showURL{%
\tempurl}


\bibitem[\protect\citeauthoryear{Maitama, Idris, and Zakari}{Maitama
  et~al\mbox{.}}{2020}]%
        {Maitama.2020}
\bibfield{author}{\bibinfo{person}{Jaafar~Zubairu Maitama},
  \bibinfo{person}{Norisma Idris}, {and} \bibinfo{person}{Abubakar Zakari}.}
  \bibinfo{year}{2020}\natexlab{}.
\newblock \showarticletitle{A Systematic Mapping Study of the Empirical
  Explicit Aspect Extractions in Sentiment Analysis}.
\newblock \bibinfo{journal}{\emph{IEEE Access}}  \bibinfo{volume}{8}
  (\bibinfo{year}{2020}), \bibinfo{pages}{113878--113899}.
\newblock
\urldef\tempurl%
\url{https://doi.org/10.1109/ACCESS.2020.3003625}
\showDOI{\tempurl}


\bibitem[\protect\citeauthoryear{Murgia, Ortu, Tourani, Adams, and
  Demeyer}{Murgia et~al\mbox{.}}{2018}]%
        {Murgia.2018}
\bibfield{author}{\bibinfo{person}{Alessandro Murgia}, \bibinfo{person}{Marco
  Ortu}, \bibinfo{person}{Parastou Tourani}, \bibinfo{person}{Bram Adams},
  {and} \bibinfo{person}{Serge Demeyer}.} \bibinfo{year}{2018}\natexlab{}.
\newblock \showarticletitle{An exploratory qualitative and quantitative
  analysis of emotions in issue report comments of open source systems}.
\newblock \bibinfo{journal}{\emph{Empirical Software Engineering}}
  \bibinfo{volume}{23}, \bibinfo{number}{1} (\bibinfo{year}{2018}),
  \bibinfo{pages}{521--564}.
\newblock
\showISSN{1382-3256}
\urldef\tempurl%
\url{https://doi.org/10.1007/s10664-017-9526-0}
\showDOI{\tempurl}


\bibitem[\protect\citeauthoryear{Murgia, Tourani, Adams, and Ortu}{Murgia
  et~al\mbox{.}}{2014}]%
        {Murgia.2014}
\bibfield{author}{\bibinfo{person}{Alessandro Murgia},
  \bibinfo{person}{Parastou Tourani}, \bibinfo{person}{Bram Adams}, {and}
  \bibinfo{person}{Marco Ortu}.} \bibinfo{year}{2014}\natexlab{}.
\newblock \showarticletitle{Do developers feel emotions? an exploratory
  analysis of emotions in software artifacts}. In
  \bibinfo{booktitle}{\emph{11th Working Conference on Mining Software
  Repositories : proceedings : May 31 - June 1, 2014, Hyderabad, India}},
  \bibfield{editor}{\bibinfo{person}{Sung Kim}, \bibinfo{person}{Martin
  Pinzger}, {and} \bibinfo{person}{Premkumar Devanbu}} (Eds.).
  \bibinfo{publisher}{ACM}, \bibinfo{address}{[Place of publication not
  identified]}.
\newblock
\showISBNx{9781450328630}
\urldef\tempurl%
\url{https://doi.org/10.1145/2597073.2597086}
\showDOI{\tempurl}


\bibitem[\protect\citeauthoryear{Müller and Fritz}{Müller and Fritz}{2015}]%
        {7194617}
\bibfield{author}{\bibinfo{person}{Sebastian~C. Müller} {and}
  \bibinfo{person}{Thomas Fritz}.} \bibinfo{year}{2015}\natexlab{}.
\newblock \showarticletitle{Stuck and Frustrated or in Flow and Happy: Sensing
  Developers' Emotions and Progress}. In \bibinfo{booktitle}{\emph{2015
  IEEE/ACM 37th IEEE International Conference on Software Engineering}},
  Vol.~\bibinfo{volume}{1}. \bibinfo{pages}{688--699}.
\newblock
\urldef\tempurl%
\url{https://doi.org/10.1109/ICSE.2015.334}
\showDOI{\tempurl}


\bibitem[\protect\citeauthoryear{Nayebi, Farahi, and Ruhe}{Nayebi
  et~al\mbox{.}}{2017}]%
        {Nayebi.2017}
\bibfield{author}{\bibinfo{person}{Maleknaz Nayebi}, \bibinfo{person}{Homayoon
  Farahi}, {and} \bibinfo{person}{Guenther Ruhe}.}
  \bibinfo{year}{2017}\natexlab{}.
\newblock \showarticletitle{Which Version Should Be Released to App Store?}. In
  \bibinfo{booktitle}{\emph{11th ACM/IEEE International Symposium on Empirical
  Software Engineering and Measurement}}. \bibinfo{publisher}{IEEE},
  \bibinfo{address}{Piscataway, NJ}, \bibinfo{pages}{324--333}.
\newblock
\showISBNx{9781509040391}
\urldef\tempurl%
\url{https://doi.org/10.1109/ESEM.2017.46}
\showDOI{\tempurl}


\bibitem[\protect\citeauthoryear{{Nicole Novielli}, {Daniela Girardi}, and
  {Filippo Lanubile}}{{Nicole Novielli} et~al\mbox{.}}{2018}]%
        {N.Novielli.2018}
\bibfield{author}{\bibinfo{person}{{Nicole Novielli}},
  \bibinfo{person}{{Daniela Girardi}}, {and} \bibinfo{person}{{Filippo
  Lanubile}}.} \bibinfo{year}{2018}\natexlab{}.
\newblock \showarticletitle{A Benchmark Study on Sentiment Analysis for
  Software Engineering Research}. In \bibinfo{booktitle}{\emph{2018 IEEE/ACM
  15th International Conference on Mining Software Repositories (MSR)}}.
  \bibinfo{pages}{364--375}.
\newblock
\showISBNx{2574-3864}


\bibitem[\protect\citeauthoryear{Niinimaki, Piri, and Lassenius}{Niinimaki
  et~al\mbox{.}}{2009}]%
        {5196929}
\bibfield{author}{\bibinfo{person}{Tuomas Niinimaki}, \bibinfo{person}{Arttu
  Piri}, {and} \bibinfo{person}{Casper Lassenius}.}
  \bibinfo{year}{2009}\natexlab{}.
\newblock \showarticletitle{Factors Affecting Audio and Text-Based
  Communication Media Choice in Global Software Development Projects}. In
  \bibinfo{booktitle}{\emph{2009 Fourth IEEE International Conference on Global
  Software Engineering}}. \bibinfo{pages}{153--162}.
\newblock
\urldef\tempurl%
\url{https://doi.org/10.1109/ICGSE.2009.23}
\showDOI{\tempurl}


\bibitem[\protect\citeauthoryear{Novielli, Calefato, Dongiovanni, Girardi, and
  Lanubile}{Novielli et~al\mbox{.}}{2020a}]%
        {Novielli.}
\bibfield{author}{\bibinfo{person}{Nicole Novielli}, \bibinfo{person}{Fabio
  Calefato}, \bibinfo{person}{Davide Dongiovanni}, \bibinfo{person}{Daniela
  Girardi}, {and} \bibinfo{person}{Filippo Lanubile}.}
  \bibinfo{year}{2020}\natexlab{a}.
\newblock \showarticletitle{Can We Use SE-specific Sentiment Analysis Tools in
  a Cross-Platform Setting?}
\newblock \bibinfo{journal}{\emph{Proceedings of the 17th International
  Conference on Mining Software Repositories}} (\bibinfo{date}{Jun}
  \bibinfo{year}{2020}).
\newblock
\showISBNx{9781450375177}
\urldef\tempurl%
\url{https://doi.org/10.1145/3379597.3387446}
\showDOI{\tempurl}


\bibitem[\protect\citeauthoryear{Novielli, Calefato, Dongiovanni, Girardi, and
  Lanubile}{Novielli et~al\mbox{.}}{2020b}]%
        {.2020}
\bibfield{author}{\bibinfo{person}{Nicole Novielli}, \bibinfo{person}{Fabio
  Calefato}, \bibinfo{person}{Davide Dongiovanni}, \bibinfo{person}{Daniela
  Girardi}, {and} \bibinfo{person}{Filippo Lanubile}.}
  \bibinfo{year}{2020}\natexlab{b}.
\newblock \bibinfo{title}{A gold standard for polarity of emotions of software
  developers in GitHub}.
\newblock
\newblock
\urldef\tempurl%
\url{https://doi.org/10.6084/m9.figshare.11604597}
\showDOI{\tempurl}


\bibitem[\protect\citeauthoryear{Novielli, Calefato, and Lanubile}{Novielli
  et~al\mbox{.}}{2014}]%
        {10.1145/2661685.2661689}
\bibfield{author}{\bibinfo{person}{Nicole Novielli}, \bibinfo{person}{Fabio
  Calefato}, {and} \bibinfo{person}{Filippo Lanubile}.}
  \bibinfo{year}{2014}\natexlab{}.
\newblock \showarticletitle{Towards Discovering the Role of Emotions in Stack
  Overflow}. In \bibinfo{booktitle}{\emph{Proceedings of the 6th International
  Workshop on Social Software Engineering}} (Hong Kong, China)
  \emph{(\bibinfo{series}{SSE 2014})}. \bibinfo{publisher}{Association for
  Computing Machinery}, \bibinfo{address}{New York, NY, USA},
  \bibinfo{pages}{33–36}.
\newblock
\showISBNx{9781450332279}
\urldef\tempurl%
\url{https://doi.org/10.1145/2661685.2661689}
\showURL{%
\tempurl}


\bibitem[\protect\citeauthoryear{Novielli, Calefato, and Lanubile}{Novielli
  et~al\mbox{.}}{2015}]%
        {Novielli.2015}
\bibfield{author}{\bibinfo{person}{Nicole Novielli}, \bibinfo{person}{Fabio
  Calefato}, {and} \bibinfo{person}{Filippo Lanubile}.}
  \bibinfo{year}{2015}\natexlab{}.
\newblock \showarticletitle{The Challenges of Sentiment Detection in the Social
  Programmer Ecosystem}. In \bibinfo{booktitle}{\emph{Proceedings of the 7th
  International Workshop on Social Software Engineering - SSE 2015}}
  \emph{(\bibinfo{series}{SSE 2015})}, \bibfield{editor}{\bibinfo{person}{Imed
  Hammouda} {and} \bibinfo{person}{Alberto Sillitti}} (Eds.).
  \bibinfo{publisher}{{ACM Press}}, \bibinfo{address}{New York, New York, USA},
  \bibinfo{pages}{33--40}.
\newblock
\showISBNx{9781450338189}
\urldef\tempurl%
\url{https://doi.org/10.1145/2804381.2804387}
\showDOI{\tempurl}


\bibitem[\protect\citeauthoryear{Novielli, Calefato, Lanubile, and
  Serebrenik}{Novielli et~al\mbox{.}}{2020c}]%
        {Novielli.20.10.2020}
\bibfield{author}{\bibinfo{person}{Nicole Novielli}, \bibinfo{person}{Fabio
  Calefato}, \bibinfo{person}{Filippo Lanubile}, {and}
  \bibinfo{person}{Alexander Serebrenik}.} \bibinfo{year}{2020}\natexlab{c}.
\newblock \bibinfo{title}{Assessment of SE-specific Sentiment Analysis Tools:
  An Extended Replication Study}.
\newblock
\newblock
\urldef\tempurl%
\url{https://doi.org/10.1007/s10664-021-09960-w}
\showDOI{\tempurl}


\bibitem[\protect\citeauthoryear{Obaidi and Klünder}{Obaidi and
  Klünder}{2021}]%
        {martin_obaidi_2021_4726651}
\bibfield{author}{\bibinfo{person}{Martin Obaidi} {and} \bibinfo{person}{Jil
  Klünder}.} \bibinfo{year}{2021}\natexlab{}.
\newblock \bibinfo{booktitle}{\emph{{Dataset: Systematic Literature Review on
  the Development and Application of Sentiment Analysis Tools in Software
  Engineering}}}.
\newblock
\urldef\tempurl%
\url{https://doi.org/10.5281/zenodo.4726651}
\showDOI{\tempurl}


\bibitem[\protect\citeauthoryear{Panichella, Di~Sorbo, Guzman, Visaggio,
  Canfora, and Gall}{Panichella et~al\mbox{.}}{2015}]%
        {S.Panichella.2015}
\bibfield{author}{\bibinfo{person}{Sebastiano Panichella},
  \bibinfo{person}{Andrea Di~Sorbo}, \bibinfo{person}{Emitza Guzman},
  \bibinfo{person}{Corrado~A. Visaggio}, \bibinfo{person}{Gerardo Canfora},
  {and} \bibinfo{person}{Harald~C. Gall}.} \bibinfo{year}{2015}\natexlab{}.
\newblock \showarticletitle{How can i improve my app? Classifying user reviews
  for software maintenance and evolution}. In \bibinfo{booktitle}{\emph{2015
  IEEE International Conference on Software Maintenance and Evolution
  (ICSME)}}. \bibinfo{pages}{281--290}.
\newblock
\urldef\tempurl%
\url{https://doi.org/10.1109/ICSM.2015.7332474}
\showDOI{\tempurl}


\bibitem[\protect\citeauthoryear{{Perry}, {Staudenmayer}, and {Votta}}{{Perry}
  et~al\mbox{.}}{1994}]%
        {300082}
\bibfield{author}{\bibinfo{person}{D.~E. {Perry}}, \bibinfo{person}{N.~A.
  {Staudenmayer}}, {and} \bibinfo{person}{L.~G. {Votta}}.}
  \bibinfo{year}{1994}\natexlab{}.
\newblock \showarticletitle{People, organizations, and process improvement}.
\newblock \bibinfo{journal}{\emph{IEEE Software}} \bibinfo{volume}{11},
  \bibinfo{number}{4} (\bibinfo{year}{1994}), \bibinfo{pages}{36--45}.
\newblock
\urldef\tempurl%
\url{https://doi.org/10.1109/52.300082}
\showDOI{\tempurl}


\bibitem[\protect\citeauthoryear{Petersen, Feldt, Mujtaba, and
  Mattsson}{Petersen et~al\mbox{.}}{2008}]%
        {Petersen.2008}
\bibfield{author}{\bibinfo{person}{Kai Petersen}, \bibinfo{person}{Robert
  Feldt}, \bibinfo{person}{Shahid Mujtaba}, {and} \bibinfo{person}{Michael
  Mattsson}.} \bibinfo{year}{2008}\natexlab{}.
\newblock \showarticletitle{Systematic Mapping Studies in Software
  Engineering}. \bibinfo{publisher}{{BCS Learning {\&} Development}}.
\newblock
\urldef\tempurl%
\url{https://doi.org/10.14236/ewic/ease2008.8}
\showDOI{\tempurl}


\bibitem[\protect\citeauthoryear{Plutchik}{Plutchik}{1980}]%
        {PLUTCHIK19803}
\bibfield{author}{\bibinfo{person}{Robert Plutchik}.}
  \bibinfo{year}{1980}\natexlab{}.
\newblock \showarticletitle{Chapter 1 - A GENERAL PSYCHOEVOLUTIONARY THEORY OF
  EMOTION}.
\newblock In \bibinfo{booktitle}{\emph{Theories of Emotion}},
  \bibfield{editor}{\bibinfo{person}{Robert Plutchik} {and}
  \bibinfo{person}{Henry Kellerman}} (Eds.). \bibinfo{publisher}{Academic
  Press}, \bibinfo{pages}{3--33}.
\newblock
\showISBNx{978-0-12-558701-3}
\urldef\tempurl%
\url{https://doi.org/10.1016/B978-0-12-558701-3.50007-7}
\showDOI{\tempurl}


\bibitem[\protect\citeauthoryear{Prenner, Unger-Windeler, and
  Schneider}{Prenner et~al\mbox{.}}{2020}]%
        {10.1145/3379177.3388907}
\bibfield{author}{\bibinfo{person}{Nils Prenner}, \bibinfo{person}{Carolin
  Unger-Windeler}, {and} \bibinfo{person}{Kurt Schneider}.}
  \bibinfo{year}{2020}\natexlab{}.
\newblock \showarticletitle{How Are Hybrid Development Approaches Organized? A
  Systematic Literature Review}. In \bibinfo{booktitle}{\emph{Proceedings of
  the International Conference on Software and System Processes}} (Seoul,
  Republic of Korea) \emph{(\bibinfo{series}{ICSSP '20})}.
  \bibinfo{publisher}{Association for Computing Machinery},
  \bibinfo{address}{New York, NY, USA}, \bibinfo{pages}{145–154}.
\newblock
\showISBNx{9781450375122}
\urldef\tempurl%
\url{https://doi.org/10.1145/3379177.3388907}
\showDOI{\tempurl}


\bibitem[\protect\citeauthoryear{Schneider, Klünder, Kortum, Handke, Straube,
  and Kauffeld}{Schneider et~al\mbox{.}}{2018}]%
        {SCHNEIDER201859}
\bibfield{author}{\bibinfo{person}{Kurt Schneider}, \bibinfo{person}{Jil
  Klünder}, \bibinfo{person}{Fabian Kortum}, \bibinfo{person}{Lisa Handke},
  \bibinfo{person}{Julia Straube}, {and} \bibinfo{person}{Simone Kauffeld}.}
  \bibinfo{year}{2018}\natexlab{}.
\newblock \showarticletitle{Positive affect through interactions in meetings:
  The role of proactive and supportive statements}.
\newblock \bibinfo{journal}{\emph{Journal of Systems and Software}}
  \bibinfo{volume}{143} (\bibinfo{year}{2018}), \bibinfo{pages}{59 -- 70}.
\newblock
\showISSN{0164-1212}
\urldef\tempurl%
\url{https://doi.org/10.1016/j.jss.2018.05.001}
\showDOI{\tempurl}


\bibitem[\protect\citeauthoryear{Shevtsov, Berekmeri, Weyns, and
  Maggio}{Shevtsov et~al\mbox{.}}{2018}]%
        {7929422}
\bibfield{author}{\bibinfo{person}{Stepan Shevtsov}, \bibinfo{person}{Mihaly
  Berekmeri}, \bibinfo{person}{Danny Weyns}, {and} \bibinfo{person}{Martina
  Maggio}.} \bibinfo{year}{2018}\natexlab{}.
\newblock \showarticletitle{Control-Theoretical Software Adaptation: A
  Systematic Literature Review}.
\newblock \bibinfo{journal}{\emph{IEEE Transactions on Software Engineering}}
  \bibinfo{volume}{44}, \bibinfo{number}{8} (\bibinfo{year}{2018}),
  \bibinfo{pages}{784--810}.
\newblock
\urldef\tempurl%
\url{https://doi.org/10.1109/TSE.2017.2704579}
\showDOI{\tempurl}


\bibitem[\protect\citeauthoryear{Storey, Treude, van Deursen, and Cheng}{Storey
  et~al\mbox{.}}{2010}]%
        {10.1145/1882362.1882435}
\bibfield{author}{\bibinfo{person}{Margaret-Anne Storey},
  \bibinfo{person}{Christoph Treude}, \bibinfo{person}{Arie van Deursen}, {and}
  \bibinfo{person}{Li-Te Cheng}.} \bibinfo{year}{2010}\natexlab{}.
\newblock \showarticletitle{The Impact of Social Media on Software Engineering
  Practices and Tools}. In \bibinfo{booktitle}{\emph{Proceedings of the FSE/SDP
  Workshop on Future of Software Engineering Research}} (Santa Fe, New Mexico,
  USA) \emph{(\bibinfo{series}{FoSER '10})}. \bibinfo{publisher}{Association
  for Computing Machinery}, \bibinfo{address}{New York, NY, USA},
  \bibinfo{pages}{359–364}.
\newblock
\showISBNx{9781450304276}
\urldef\tempurl%
\url{https://doi.org/10.1145/1882362.1882435}
\showDOI{\tempurl}


\bibitem[\protect\citeauthoryear{Thelwall, Buckley, and Paltoglou}{Thelwall
  et~al\mbox{.}}{2012}]%
        {Thelwall.2012}
\bibfield{author}{\bibinfo{person}{Mike Thelwall}, \bibinfo{person}{Kevan
  Buckley}, {and} \bibinfo{person}{Georgios Paltoglou}.}
  \bibinfo{year}{2012}\natexlab{}.
\newblock \showarticletitle{Sentiment strength detection for the social web}.
\newblock \bibinfo{journal}{\emph{Journal of the American Society for
  Information Science and Technology}} \bibinfo{volume}{63},
  \bibinfo{number}{1} (\bibinfo{year}{2012}), \bibinfo{pages}{163--173}.
\newblock
\showISSN{1532-2882}
\urldef\tempurl%
\url{https://doi.org/10.1002/asi.21662}
\showDOI{\tempurl}


\bibitem[\protect\citeauthoryear{Thelwall, Buckley, Paltoglou, {Di Cai}, and
  Kappas}{Thelwall et~al\mbox{.}}{2010}]%
        {Thelwall.2010}
\bibfield{author}{\bibinfo{person}{Mike Thelwall}, \bibinfo{person}{Kevan
  Buckley}, \bibinfo{person}{Georgios Paltoglou}, \bibinfo{person}{{Di Cai}},
  {and} \bibinfo{person}{Arvid Kappas}.} \bibinfo{year}{2010}\natexlab{}.
\newblock \showarticletitle{Sentiment strength detection in short informal
  text}.
\newblock \bibinfo{journal}{\emph{Journal of the American Society for
  Information Science and Technology}} \bibinfo{volume}{61},
  \bibinfo{number}{12} (\bibinfo{year}{2010}), \bibinfo{pages}{2544--2558}.
\newblock
\showISSN{1532-2882}
\urldef\tempurl%
\url{https://doi.org/10.1002/asi.21416}
\showDOI{\tempurl}


\bibitem[\protect\citeauthoryear{Tourani, Jiang, and Adams}{Tourani
  et~al\mbox{.}}{2014}]%
        {10.5555/2735522.2735528}
\bibfield{author}{\bibinfo{person}{Parastou Tourani}, \bibinfo{person}{Yujuan
  Jiang}, {and} \bibinfo{person}{Bram Adams}.} \bibinfo{year}{2014}\natexlab{}.
\newblock \showarticletitle{Monitoring Sentiment in Open Source Mailing Lists:
  Exploratory Study on the Apache Ecosystem}. In
  \bibinfo{booktitle}{\emph{Proceedings of 24th Annual International Conference
  on Computer Science and Software Engineering}} (Markham, Ontario, Canada)
  \emph{(\bibinfo{series}{CASCON '14})}. \bibinfo{publisher}{IBM Corp.},
  \bibinfo{address}{USA}, \bibinfo{pages}{34–44}.
\newblock


\bibitem[\protect\citeauthoryear{Umer, Liu, and Illahi}{Umer
  et~al\mbox{.}}{2020}]%
        {Umer.2020}
\bibfield{author}{\bibinfo{person}{Qasim Umer}, \bibinfo{person}{Hui Liu},
  {and} \bibinfo{person}{Inam Illahi}.} \bibinfo{year}{2020}\natexlab{}.
\newblock \showarticletitle{CNN-Based Automatic Prioritization of Bug Reports}.
\newblock \bibinfo{journal}{\emph{IEEE Transactions on Reliability}}
  (\bibinfo{year}{2020}), \bibinfo{pages}{1--14}.
\newblock
\showISSN{0018-9529}
\urldef\tempurl%
\url{https://doi.org/10.1109/TR.2019.2959624}
\showDOI{\tempurl}


\bibitem[\protect\citeauthoryear{Watson, Clark, and Tellegen}{Watson
  et~al\mbox{.}}{1988}]%
        {Watson.1988}
\bibfield{author}{\bibinfo{person}{David Watson}, \bibinfo{person}{Lee Clark},
  {and} \bibinfo{person}{Auke Tellegen}.} \bibinfo{year}{1988}\natexlab{}.
\newblock \showarticletitle{Development and validation of brief measures of
  positive and negative affect: the PANAS scales}.
\newblock \bibinfo{journal}{\emph{Journal of personality and social
  psychology}} \bibinfo{volume}{54}, \bibinfo{number}{6}
  (\bibinfo{year}{1988}), \bibinfo{pages}{1063--1070}.
\newblock
\showISSN{0022-3514}
\urldef\tempurl%
\url{https://doi.org/10.1037//0022-3514.54.6.1063}
\showDOI{\tempurl}


\bibitem[\protect\citeauthoryear{Werner, Tapuc, Montgomery, Sharma, Dodos, and
  Damian}{Werner et~al\mbox{.}}{2018}]%
        {Werner.2018}
\bibfield{author}{\bibinfo{person}{Colin Werner}, \bibinfo{person}{Gabriel
  Tapuc}, \bibinfo{person}{Lloyd Montgomery}, \bibinfo{person}{Diksha Sharma},
  \bibinfo{person}{Sanja Dodos}, {and} \bibinfo{person}{Daniela Damian}.}
  \bibinfo{year}{2018}\natexlab{}.
\newblock \showarticletitle{How Angry are Your Customers? Sentiment Analysis of
  Support Tickets that Escalate}. In \bibinfo{booktitle}{\emph{2018 1st
  International Workshop on Affective Computing for Requirements Engineering}},
  \bibfield{editor}{\bibinfo{person}{Davide Fucci}, \bibinfo{person}{Nicole
  Novielli}, {and} \bibinfo{person}{Emitz{\'a} Guzm{\'a}n}} (Eds.).
  \bibinfo{publisher}{IEEE}, \bibinfo{address}{Piscataway, NJ}.
\newblock
\showISBNx{9781538683613}
\urldef\tempurl%
\url{https://doi.org/10.1109/affectre.2018.00006}
\showDOI{\tempurl}


\bibitem[\protect\citeauthoryear{Whiting, Gao, Xing, Diarrassouba, Nguyen, and
  Bernstein}{Whiting et~al\mbox{.}}{2020}]%
        {10.1145/3392877}
\bibfield{author}{\bibinfo{person}{Mark~E. Whiting}, \bibinfo{person}{Irena
  Gao}, \bibinfo{person}{Michelle Xing}, \bibinfo{person}{N'godjigui~Junior
  Diarrassouba}, \bibinfo{person}{Tonya Nguyen}, {and}
  \bibinfo{person}{Michael~S. Bernstein}.} \bibinfo{year}{2020}\natexlab{}.
\newblock \showarticletitle{Parallel Worlds: Repeated Initializations of the
  Same Team to Improve Team Viability}.
\newblock  \bibinfo{volume}{4}, \bibinfo{number}{CSCW1}, Article
  \bibinfo{articleno}{067} (\bibinfo{date}{May} \bibinfo{year}{2020}),
  \bibinfo{numpages}{22}~pages.
\newblock
\urldef\tempurl%
\url{https://doi.org/10.1145/3392877}
\showDOI{\tempurl}


\bibitem[\protect\citeauthoryear{Wohlin}{Wohlin}{2014}]%
        {Wohlin.2014}
\bibfield{author}{\bibinfo{person}{Claes Wohlin}.}
  \bibinfo{year}{2014}\natexlab{}.
\newblock \showarticletitle{Guidelines for Snowballing in Systematic Literature
  Studies and a Replication in Software Engineering}. In
  \bibinfo{booktitle}{\emph{Proceedings of the 18th International Conference on
  Evaluation and Assessment in Software Engineering}} (London, England, United
  Kingdom) \emph{(\bibinfo{series}{EASE '14})}. \bibinfo{publisher}{Association
  for Computing Machinery}, \bibinfo{address}{New York, NY, USA}, Article
  \bibinfo{articleno}{38}, \bibinfo{numpages}{10}~pages.
\newblock
\showISBNx{9781450324762}
\urldef\tempurl%
\url{https://doi.org/10.1145/2601248.2601268}
\showDOI{\tempurl}


\bibitem[\protect\citeauthoryear{Wohlin, Runeson, Höst, Ohlsson, Regnell, and
  Wesslén}{Wohlin et~al\mbox{.}}{2012}]%
        {Wohlin.2012}
\bibfield{author}{\bibinfo{person}{Claes Wohlin}, \bibinfo{person}{Per
  Runeson}, \bibinfo{person}{Martin Höst}, \bibinfo{person}{Magnus~C.
  Ohlsson}, \bibinfo{person}{Björn Regnell}, {and} \bibinfo{person}{Anders
  Wesslén}.} \bibinfo{year}{2012}\natexlab{}.
\newblock \bibinfo{booktitle}{\emph{Experimentation in software engineering}}.
\newblock \bibinfo{publisher}{Springer}, \bibinfo{address}{Berlin}.
\newblock
\showISBNx{9783642290442}
\urldef\tempurl%
\url{https://doi.org/10.1007/978-3-642-29044-2}
\showDOI{\tempurl}


\bibitem[\protect\citeauthoryear{Yasin, Fatima, Wen, Afzal, Azhar, and
  Torkar}{Yasin et~al\mbox{.}}{2020}]%
        {8984351}
\bibfield{author}{\bibinfo{person}{Affan Yasin}, \bibinfo{person}{Rubia
  Fatima}, \bibinfo{person}{Lijie Wen}, \bibinfo{person}{Wasif Afzal},
  \bibinfo{person}{Muhammad Azhar}, {and} \bibinfo{person}{Richard Torkar}.}
  \bibinfo{year}{2020}\natexlab{}.
\newblock \showarticletitle{On Using Grey Literature and Google Scholar in
  Systematic Literature Reviews in Software Engineering}.
\newblock \bibinfo{journal}{\emph{IEEE Access}}  \bibinfo{volume}{8}
  (\bibinfo{year}{2020}), \bibinfo{pages}{36226--36243}.
\newblock
\urldef\tempurl%
\url{https://doi.org/10.1109/ACCESS.2020.2971712}
\showDOI{\tempurl}


\bibitem[\protect\citeauthoryear{Zahra, Azam, Ilyas, Faisal, Ambreen, and
  Gondal}{Zahra et~al\mbox{.}}{2017}]%
        {10.1145/3029387.3029392}
\bibfield{author}{\bibinfo{person}{Kinza Zahra}, \bibinfo{person}{Farooque
  Azam}, \bibinfo{person}{Fauqia Ilyas}, \bibinfo{person}{Huma Faisal},
  \bibinfo{person}{Nadia Ambreen}, {and} \bibinfo{person}{Nida Gondal}.}
  \bibinfo{year}{2017}\natexlab{}.
\newblock \showarticletitle{Success Factors of Organizational Change in
  Software Process Improvement: A Systematic Literature Review}. In
  \bibinfo{booktitle}{\emph{Proceedings of the 5th International Conference on
  Information and Education Technology}} (Tokyo, Japan)
  \emph{(\bibinfo{series}{ICIET '17})}. \bibinfo{publisher}{Association for
  Computing Machinery}, \bibinfo{address}{New York, NY, USA},
  \bibinfo{pages}{155–160}.
\newblock
\showISBNx{9781450348034}
\urldef\tempurl%
\url{https://doi.org/10.1145/3029387.3029392}
\showDOI{\tempurl}


\bibitem[\protect\citeauthoryear{Zhang, Xu, Thung, Haryono, Lo, and
  Jiang}{Zhang et~al\mbox{.}}{2020}]%
        {9240704}
\bibfield{author}{\bibinfo{person}{Ting Zhang}, \bibinfo{person}{Bowen Xu},
  \bibinfo{person}{Ferdian Thung}, \bibinfo{person}{Stefanus~Agus Haryono},
  \bibinfo{person}{David Lo}, {and} \bibinfo{person}{Lingxiao Jiang}.}
  \bibinfo{year}{2020}\natexlab{}.
\newblock \showarticletitle{Sentiment Analysis for Software Engineering: How
  Far Can Pre-trained Transformer Models Go?}. In
  \bibinfo{booktitle}{\emph{2020 IEEE International Conference on Software
  Maintenance and Evolution (ICSME)}}. \bibinfo{pages}{70--80}.
\newblock
\urldef\tempurl%
\url{https://doi.org/10.1109/ICSME46990.2020.00017}
\showDOI{\tempurl}


\end{thebibliography}

%%
%% If your work has an appendix, this is the place to put it.
\appendix

\end{document}